\documentclass[12pt,preprint]{aastex}
\usepackage{epsfig}
\usepackage{graphicx}
\usepackage{color}
\usepackage{enumerate}
\usepackage{ulem}

\def\beq{\begin{equation}}
\def\eeq{\end{equation}}

\def\mbh{M_{\rm BH}}
\def\rt{r_{\rm t}}

\def\rp{r_{\rm p}}
\def\rsch{r_{\rm g}}

\def\msun{M_\odot}
\def\rsun{R_\odot}

\def\sdss{SDSS~J1201+30~}

\shorttitle{A dormant SMBHB in SDSS J120136.02+300305.5}
\shortauthors{Liu et al.}

\begin{document}

\title{A milliparsec supermassive black hole binary candidate
  in the galaxy SDSS J120136.02+300305.5}


\author{F.K. Liu\altaffilmark{1,2}, Shuo Li\altaffilmark{3,1}, and
  S. Komossa\altaffilmark{4,5}}

\altaffiltext{1}{Department of Astronomy, Peking University, Beijing
  100871, China; {\it fkliu@pku.edu.cn}}
\altaffiltext{2}{Kavli Institute for Astronomy and Astrophysics,
  Peking University, Beijing 100871, China}
\altaffiltext{3}{National Astronomical Observatories, Chinese
  Academy of Sciences, 100012 Beijing, China}
\altaffiltext{4}{Max-Planck-Institut f\"ur Radioastronomie, Auf dem
  H\"ugel 69, 53121 Bonn, Germany}
\altaffiltext{5}{Kavli Institute for Theoretical Physics, Santa
  Barbara, CA 93106, USA}

\begin{abstract}
Galaxy mergers play a key role in the evolution of galaxies and the
growth of their central supermassive black holes (SMBHs).
A search for (active) SMBH binaries (SMBHBs)
at the centers of the merger remnants is currently ongoing. Perhaps
the greatest challenge is to identify the {\bf \it inactive}
SMBHBs, which might be the most abundant, but are also the most difficult
to identify. Liu et al. predicted characteristic drops in the
light curves of tidal disruption events (TDEs), caused by the presence
of a secondary SMBH. Here, we apply that model to the light curve
of the optically inactive galaxy SDSS J120136.02+300305.5, which was
identified as a candidate TDE with {\it XMM-Newton}. We show that the deep
dips in its evolving X-ray light curve can be well explained by the
presence of a SMBHB at its core. A SMBHB model with a mass of the
primary of $M_{\rm BH} = 10^7 M_\sun$, a mass ratio $q \simeq 0.08$,
and a semimajor axis $a_{\rm b} \simeq 0.6\, {\rm mpc}$ is in good
agreement with the observations. Given that primary mass, introducing
an orbital eccentricity is needed, with $e_{\rm b} \simeq
0.3$. Alternatively, a lower mass primary of $M_{\rm BH} = 10^6
M_\sun$ in a circular orbit fits the light curve well. Tight binaries
like this one, which have already overcome the ``final parsec problem,'' 
are prime sources of gravitational wave radiation once the two SMBHs 
coalesce. Future transient surveys, which will detect TDEs in large
numbers, will place tight constraints on the SMBHB fraction in
otherwise non-active galaxies.
\end{abstract}


\section{Introduction}
\label{introduction}

During the hierarchical formation of galaxies, merging galaxies
quickly bring two supermassive black holes (SMBHs) into their
center, forming a hard SMBH binary (SMBHB) at parsec
scale \citep{beg80,qui96,vol03,may07,tan09,kul12,kul13}. Further
evolution of hard SMBHBs
depends very much on the detailed nuclear structures. A hard SMBHB may
stall at the parsec scale for a timescale longer than the Hubble time
in the spherical and isotropic galactic nuclei \citep{beg80}, but can
be driven to the
strong gravitational wave (GW) regime at milliparsec scale by efficient
stellar dynamical processes in non-spherical, anisotropic, and/or rotating
nuclear clusters \citep{mer04,ber06,pre07,pre11}, or by
hydrodynamic processes in massive gas disks \citep[][and references
therein]{gou00,koc12,col11}. If the orbits of SMBHBs could be highly
eccentric \citep{arm05,ber06,pre11,iwa11,ses11,che11}, the strong
GW regime may extend far out and the SMBHBs 
can quickly evolve to the GW regime. The SMBHBs in the strong
GW regime coalesce within a Hubble time. They are the
primary targets for the proposed {\it Laser Interferometer Space Antenna}
({\it eLISA}) and ongoing Pulsar Timing Arrays (PTAs).

A few observations of SMBHBs in gaseous systems have been reported,
ranging from spatially resolved binary active galactic nuclei
\citep[AGNs; e.g.,][]{kom03,rod06,gre10,liu10,fab11}, to candidate
un-resolved binary systems. The observational evidence for the latter
is based on double-peaked broad emission lines in
quasars and their variability \citep[e.g.,][]{tsa11,ju13,she13};
characteristic spatial structures in radio jets
\citep[e.g.,][]{beg80,con95,liu07,rol08,bri12}; quasi-periodic
outbursts in some blazars \citep[e.g.,][and references
  therein]{sil88,liu95,liu97,liu06,liu02,qia07,val11}; and swift jet
reorientation in
X-shaped radio galaxies \citep{liu04}. Further, some candidate systems
of coalesced SMBHBs have been identified, based on double-double radio
galaxies that show an interruption and recurrence of jet formation
\citep{liu03}, or based on evidence for recoiled SMBHs
\citep[e.g.,][]{kom08,liu12a,civ12}. All the observational
evidence for SMBHBs in AGNs is consistent with the
scenario of rapid migration of SMBHBs within massive gas disks.

However, most SMBHBs may form quiescently either in gas-poor or
minor galaxy mergers without driving AGN activities.
The high frequency of minor and gas-poor galaxy mergers is
important in the
formation and evolution of both late type and massive elliptical
galaxies \citep[e.g.,][]{dok10,naa09,mcw12}, therefore it is essential
to understand the evolution of SMBHBs in quiescent galaxies, not only
for the GW detections with {\it LISA} and PTA, but also to 
test the galaxy formation and evolution models. It is a challenge to
detect dormant SMBHBs and observationally constrain their evolution in
galactic nuclei. A dormant SMBH can be investigated observationally
when it tidally disrupts stars passing by and becomes a transient
AGN \citep{hil75,lum85,ree88,phi89,lod09}. More than 20 candidates
for stellar tidal disruptions (TDEs) by SMBHs have been reported
in wavebands from $\gamma$-ray and X-ray through UV and optical to radio
frequencies
(\citep[e.g.,][]{kom99,kom02,kom08b,gez08,gez12,zau11,blo11,bur11,sax12};
\citep[see][for a review]{kom12}).
Recently, it was suggested that dormant SMBHBs in non-active
galaxies could also be investigated when one of the SMBHs tidally
disrupted a star, because SMBHBs can give some key imprints on the
event rates, locations, and the environment in the galactic nuclei,
light curves, and spectra of tidal flares
\citep{iva05,kom08c,che08,che09,che11,liu09,weg11,sto11,sto12a,li12,che13,liu12}.
SMBHBs can
dramatically change the tidal disruption rates of stars in galactic
nuclei. Our investigations showed that stellar tidal disruption rates
by ultra-compact SMBHBs, particularly in the GW dominated
  regime, are at least one order of magnitude lower 
\citep{che08}, while those by bound (but not-hard) SMBHBs
\citep{che09,che11} or dual SMBHs in merging galaxies \citep{liu12}
can be up to three orders of magnitudes higher than the estimated
tidal disruption rates by single SMBHs in isolated galaxies
\citep{wan04}. However, despite the significant difference in
the tidal disruption rates, the separate contributions of stellar
tidal disruptions by single, bound and hard binaries, and dual
SMBHs to the total stellar disruption events during cosmic time are
comparable to one another \citep{liu12}, implying that some of the
observed tidal disruption flares are probably from SMBHB systems.
Numerical and analytic computations indicated that the stellar
tidal disruption flares in SMBHB systems would show temporary
interruptions and recurrences. The interruptions are due to the
perturbation of the companion SMBH on the plasma streams
of the tidally disrupted star \citep[][LLC09
hereafter]{liu09}. This is one of the key signatures for SMBHBs in
galactic nuclei.

In this paper, we report the first candidate for a SMBHB in
a quiescent galaxy. The galaxy SDSS J120136.02+300305.5 (SDSS J1201+30 
hereafter) 
at redshift $z=0.146$ was observed to be at outburst in X-rays
with {\it XMM-Newton}, during the slew-survey observation on 2010
June 10, most probably due to the tidal disruption of a star 
by a SMBH at its center \citep[][S2012
hereafter]{sax12}. Follow-up observations in X-rays with the {\it
Swift} and {\it XMM-Newton} space telescopes made by
S2012 showed that the X-ray flux of the flare is, on
  average, consistent with a $t^{-5/3}$ power law, as predicted by the
canonical fallback model for tidal disruption
\citep{ree88,eva89}. Superposed on this overall decline are
  rapid, strong dips in the X-ray flux. No absorption and no radio
jet was detected in the observations. In particular, the light
curve showed that SDSS J1201+30 decayed in the X-ray flux by more than
about 50 times within seven days and became completely invisible to
{\it Swift} between 27 and 48 days after discovery. Then, it recurred
to follow the original power-law decay from 2010 October 24 to
December 23. Here, we will show that all the striking features in
the light curve of the event are challenging to understand by
the TDE model in the presence of a single SMBH, but are fully
consistent with the key predictions of the model for stellar tidal
disruption in a SMBHB system, given by LLC09. Applying the binary black
hole model for tidal disruption to SDSS J1201+30 and using the
observations, we constrain well the parameters of the SMBHB
system despite the gaps in the light curve coverage.

This paper is organized as follows. In Section~\ref{obs}, we introduce the
fallback model for tidal disruptions of stars by SMBHs and give
some general constraints on the model parameters for SDSS J1201+30,
based on the observations given by S2012. In Section~\ref{bin},
we briefly describe the model for stellar tidal disruption in a
SMBHB
system (Section~\ref{intr:an}) and present the numerical results of
employing the binary black hole model to SDSS J1201+30 in
Section~\ref{mod}. We show that the observed light curve in X-rays
can be well reproduced by a simple SMBHB model. In particular,
the question of
whether SMBHB orbits show eccentricity is important, both for
the evolution of SMBHBs in galactic nuclei and the detection of
GW radiation. We explore the effect of the orbital eccentricity of
SMBHBs on the light curves, and show that the orbit of the SMBHB in SDSS
J1201+30 must be elliptical, if the central black hole has a mass $\sim
10^7 M_\sun$ in Section~\ref{ecc}. After critically investigating the 
alternatives in Section~\ref{sec:alt}, we provide our discussion of
the orbital parameters of the SMBHB system in SDSS J1201+30 in
Section~\ref{sec:para}, the GW emission of the SMBHB system in 
Section~\ref{sec:gw}, and the frequency of SMBHBs among known TDEs
in Section~\ref{sec:freq}. Throughout the paper, we assume a
$\Lambda$CDM cosmology with parameters $H_0 = 70 \, {\rm km\; s^{-1}\;
  Mpc^{-1}}$, $\Omega_\Lambda = 0.73$, and $\Omega_M = 0.27$.

%%%%%%%%%%%%%%%

\section{Stellar tidal disruption in SDSS J1201+30}
\label{obs}

A star of mass $M_*$ and radius $R_*$ is tidally disrupted, if it
passes by a SMBH with mass $M_{\rm BH}$
at a distance less than the tidal radius
\begin{equation}
  \rt \simeq R_{*}(\mu^2 \mbh/M_{*})^{1/3}
  \label{rt}
\end{equation}
\citep{hil75,ree88}, where $\mu$ is of order unity. Because $\mu$
is degenerate with the other model parameters and cannot be
constrained independently, we adopt $\mu =1$ for simplicity. When
a star is tidally disrupted, the stellar plasma is
redistributed in the specific energy $E$ ranging from $E_{\rm b}-\Delta
E$ to $E_{\rm b}+\Delta E$, where $E_{\rm b}$ is the total energy of
the star before disruption and $\Delta E$ is the maximum specific energy
obtained by the stellar plasma at disruption. If the star is
bound to the SMBH, the total energy is $E_{\rm b} = - G M_{\rm BH} /(2
a_*)$ with $a_*$ the orbital semimajor axis of the bound star; while
the total energy is $E_{\rm b} \sim {1 \over 2} \sigma_*^2$ with
$\sigma_*$ the stellar velocity 
dispersion if the star is from the region around the influence radius
of the SMBH. The maximum specific energy is
\beq
\Delta E  \simeq  kG\mbh R_*/r_{\rm p}^2 ,
\label{eq:DE}
\eeq
where $G$ is the Newtonian gravitational constant, $r_{\rm p}$
is the orbital pericenter of the star, and $1 \le k \le 3$ is
a parameter depending on the spin of the
star at disruption \citep{ree88,li02}. If the tidal spin-up
of the star is negligible $k=1$, while $k=3$ for a star being
maximally spun up to break-up angular velocity. The spin-up of
the star
depends on the penetration factor $\beta \equiv \rt/r_{\rm p}$. For a
tidal disruption with $\beta \ga 1$ as in SDSS J1201+30 (see
discussions below) and because $k$ is degenerate with $\beta$, we take
$k=2.5$ as fiducial value. For a tidal disruption of
penetration factor $\beta$, Equation~(\ref{eq:DE}) becomes
\beq
\Delta E  \simeq  \frac{kG\mbh R_*}{\rt^2}\times \beta^l ,
\label{eq:DEb}
\eeq
where $l=2$. Equation~(\ref{eq:DEb}) suggests a strong dependence of
$\Delta E$ on $\beta$. Recent analytical and numerical calculations
indicated, however, that $\Delta E$ might be nearly independent of
$\beta$ with $l \approx 0$, because $\Delta E$ is
determined mostly by the conditions of the disrupted star around the
tidal radius \citep{gui12,sto12,hay13}. In order to study the effects
of different dependencies of $\Delta E$ on $\beta$, we also did
calculations for $l = 0$ (see Section~\ref{sec:para} and
Figure~\ref{fig:M6}). The results show that our conclusions do not 
change with the relationship of $\Delta E$ and $\beta$. Thus, we use
$l = 2$ as the fiducial value throughout the paper, except
when noticed otherwise.

The initial total energy of the star $E_{\rm b}$ is usually assumed
to be negligible in the literature \citep[except, e.g.,][]{hay13}
because stars approaching from the SMBH influence radius dominate
the TDEs in single SMBH systems with spherical isotropic distributions
of stars in the galactic nuclei
\citep[e.g.,][]{mag99,wan04}. However, the tidal disruptions are
overwhelmed by the bound stars inside the SMBH sphere of influence due
to either the scattering of the massive companion in SMBHB systems
\citep{che09,che11}, or the chaotic orbits of stars in the galactic nuclei
with triaxiality \citep[e.g.,][]{vas13,liu12} and thus the total (binding)
energy of stars is not necessary small. If the total energy of the bound
stars has $|E_{\rm b}| \ga \Delta E$, or an orbital semi-major axis
\begin{eqnarray}
a_* \la a_{\rm cr}\simeq 1177.6k^{-1}\beta^{-l}M_6^{-1/3}\rsch 
r_*m_*^{-2/3}\simeq 471 \left({k \over 2.5}\right)^{-1} \beta^{-l} 
M_6^{-1/3}\rsch r_*m_*^{-2/3}
\label{eq:acr}
\end{eqnarray}
(where $r_{\rm g}$ is the Schwarzschild radius of the black
hole, and $M_6 = \mbh / 10^6\msun$ with $M_\sun$ the solar mass,
$m_* = M_*/\msun$, and $r_* = R_*/\rsun$ with $\rsun$ the solar
  radius), the tidal 
disruption and the light curves of tidal flares are significantly
different from the predictions by the canonical fallback model for
unbound stars \citep{hay13}. This is then inconsistent with the
observations of SDSS J1201+30. For typical nuclear star clusters
around SMBHs in the galactic nuclei, the fraction of stars with $a_* \la
a_{\rm cr}$ is negligible.  Therefore, we assume $a_* \gg a_{\rm
cr}$ or $|E_{\rm b}| \ll \Delta E$ and neglect $|E_{\rm b}|$ in
the discussions of the TDE in SDSS J1201+30 from
now on. However, this does not imply that the star is necessary from
the influence radius of the SMBH.

The observations of the tidal flare in SDSS J1201+30 are consistent
with the predictions of the canonical fallback model for
stellar tidal disruption. In the canonical
fallback model, the stellar plasma after tidal disruption
distributes evenly in specific energy $E$ between $-\Delta E$ and
$\Delta E$. Therefore about half of the stellar plasma gains energy
and is ejected away from the system, while the other half loses energy and
becomes bound to the SMBH. The bound stellar plasma follows
Keplerian orbits with extremely high eccentricities and falls
back to the pericenter at a mass rate
\begin{eqnarray}
  \dot{M}(t) & \simeq & {M_{*} \over 3 t_{\rm min}} \left({t - T_{\rm D}
      \over t_{\rm min}}\right)^{-5/3} \, \rm{for}~\, \textit t \ge
      T_{\rm D} + t_{\rm min}
  \label{eq:Mdot}
\end{eqnarray}
\citep{ree88,eva89,lod09}, where $T_{\rm D} = 0$ is the epoch of
disruption and $t_{\rm min}$ is the return time of the most bound
stellar debris to the pericenter
\begin{eqnarray}
  t_{\rm min} & \simeq& 2 \pi G\mbh \left( 2\Delta E\right)^{-3/2}
  \simeq 41.0~{\rm days} \times
  k^{-3/2}\beta^{-3l/2}M_6^{1/2}r_*^{3/2}m_*^{-1} \nonumber\\
  &\simeq & 10.4 {\rm days} \times \left({k \over 2.5}\right)^{-3/2}
  \beta^{-3l/2}M_6^{1/2}r_*^{3/2}m_*^{-1} .
\label{eq:tmin}
\end{eqnarray}
Equation~(\ref{eq:tmin}) implies that $t_{\rm min}$
strongly depends on $\beta$, a parameter that cannot be given
a priori. We take $\beta$ as a free parameter and determine it
observationally. Thus, we consider only the tidal disruption of
solar-type stars and all the uncertainties are absorbed into
$\beta$. After the bound stellar debris falls back to pericenter, it
quickly circularizes at radius $r_{\rm c} \simeq  2 r_p$ due to
strong shocks  between incoming and outgoing stellar plasma
  streams, leading to accretion onto the SMBH \citep{ree88,phi89,ulm99}.
Numerical simulations suggest that the accretion rate may peak at
$t_{\rm peak} \sim 1.5t_{\rm min}$ instead of time $t_{\rm min}$
as in Equation~(\ref{eq:Mdot}) \citep{eva89,lod09}.
Equation~(\ref{eq:Mdot}) shows that the SMBH accretes at
rates larger than the Eddington accretion rate $\dot{M}_{\rm
Edd} = L_{\rm Edd} / 0.1 c^2$ with the Eddington luminosity $L_{\rm
Edd}= 1.25 \times 10^{44} M_6 \, {\rm erg\; s^{-1}}$, and $c$ the
speed of light, {until a time
\begin{eqnarray}
  t & > & t_{\rm Edd}  \simeq  776~{\rm days}\times k^{-3/5} \beta^{-3l/5}M_6^{-2/5}
  m_*^{1/5} r_*^{3/5} \nonumber \\
  &\simeq & 448 ~{\rm days} \times \left({k \over 2.5}\right)^{-3/5}
  \beta^{-3l/5}M_6^{-2/5} m_*^{1/5} r_*^{3/5} .
\label{eq:tedd}
\end{eqnarray}
For the SMBHs of mass $3\times 10^5 M_\odot < M_{\rm BH} < 2\times 10^7
M_\odot$ in SDSS J1201+30 (S2012), $ t_{\rm Edd} $ is on the order of
months to years, depending on the penetration factor $\beta$. After
a time
\begin{eqnarray}
  t & > & t_{\rm ADAF}  \simeq 33.7 ~{\rm yr}\times k^{-3/5}
  \beta^{-3l/5}M_6^{-2/5} \dot{m}_{-2}^{-3/5}
  m_*^{1/5} r_*^{3/5} \nonumber \\
  &\simeq & 19.5 ~{\rm yr} \times \left({k \over 2.5}\right)^{-3/5}
  \beta^{-3l/5}M_6^{-2/5} m_*^{1/5} r_*^{3/5} \dot{m}_{-2}^{-3/5}
\label{eq:adaf}
\end{eqnarray}
with $\dot{m}_{-2} = \dot{m}/ 0.01$,
the accretion rate $\dot{m} = \dot{M} /\dot{M}_{\rm Edd}$ becomes less
than 0.01 and the accretion mode probably becomes advection
dominated \citep{nar94}.} For a period of order of years, in which we
are interested here, we have $\dot{m} > 0.01$ at high-states, 
and thus do not consider the change of accretion mode from
radiatively efficient to inefficient.

To relate the fallback rates of stellar plasma to the
observed X-ray flux in the 0.2-2 keV band, we need to know
the ratio of the X-ray and bolometric luminosity, the transfer efficiency
of matter to radiation, the fraction of the advected and
photon-trapped energy into the SMBH, the fraction of the disk
outflow to the total matter, the contribution and absorption of the
disk corona, the amount of reflection, and the disk orientation.
Because most of the processes cannot be determined from first
principles, we simply adopt
\beq
L_X = f_{\rm x} \dot{M}c^2 ,
\label{eq:fl_mat}
\eeq
with $f_{\rm x}$ a free parameter to be fixed observationally. We
assume $f_{\rm x}$ to be constant for the period we are interested in,
and expect that this is a good assumption \citep[e.g.,][]{tan13}.
From Equations~(\ref{eq:Mdot}) and (\ref{eq:fl_mat}) and the data in
Table~1 of S2012, we find that the tidal disruption
happened about 11.8~days in the observer frame or $t_1 \simeq 10.3 \,
{\rm days}$ in the object rest frame before the first detection on 2010
June 10. To obtain $t_1$, we have used all the detections of
SDSS J1201+30 given in Table 1 of S2012 except the
observation on 2010 November 23 which was strongly affected by
radiation. Because $f_{\rm x}$ in Equation~(\ref{eq:fl_mat}) and
$t_{\rm min}$ in Equation~(\ref{eq:Mdot})
are degenerate and cannot be determined independently by fitting the
observational data, we can give only an upper limit to
$t_{\rm min}$ as $t_{\rm min} \le 10.3\, {\rm days}$. {If
$l=2$, from Equation~(\ref{eq:tmin}) we have 
\begin{eqnarray}
  \beta & \simeq & \left({t_{\rm min} \over 41.0 {\rm days}}\right)^{-1/3}
  \times k^{-1/2} M_6^{1/6}r_*^{1/2}m_*^{-1/3} \ga 1.0 \left({k \over
  2.5}\right)^{-1/2} M_6^{1/6}r_*^{1/2}m_*^{-1/3} .
\label{eq:beta}
\end{eqnarray}
Thus the tidal disruption of the star in SDSS J1201+30 should
occur inside the tidal radius with $\beta \ga 1$. For $l=0$, $\Delta
E$ is independent of $\beta$, which gives a constraint on $k$
alternatively
\begin{eqnarray}
  k & \simeq & \left({t_{\rm min} \over 41.0 {\rm days}}\right)^{-2/3}
  \times M_6^{1/3}r_*m_*^{-2/3} \ga 2.5 M_6^{1/3}r_*m_*^{-2/3} .
\label{eq:betak}
\end{eqnarray}

The above analyses were carried out in Newtonian gravity. However,
the disrupted star has a pericenter $r_{\rm p} = \beta^{-1}
  r_{\rm t} \simeq 12 M_6^{-2/3} (\beta/2)^{-1} r_{\rm 
g}$ and general relativity (GR) effects are important for a
SMBH with mass $M_{\rm BH} \ga 10^6 M_\odot$.
To take the GR effect into account, we use the
pseudo-Newtonian potential $\phi = G\mbh /(r - \rsch)$ \citep{pac80}
in our numerical simulations. Thus, Equation~(\ref{eq:DE}) is
rewritten
\begin{eqnarray}
\Delta E  \simeq \frac{kG\mbh}{1 -\rsch/\rp}\frac{R_*}{\rp^2} =
  \frac{k G\mbh}{1 -\beta^{l/2} \rsch/\rt} \frac{R_*}{\rt^2} \beta^l .
\label{eq:DEp}
\end{eqnarray}
With a larger maximum specific energy $\Delta E$,
Equation~(\ref{eq:tmin}) gives a smaller $t_{\rm min}$, leading to
a larger $\beta$ and stronger GR effects.}

%%%%%%%%%
\section{A SMBHB system in SDSS J1201+30}
\label{bin}

\subsection{SMBHB Model and Dynamic Evolution of the Accretion
  Disk} 
\label{intr:an}

When SDSS J1201+30 becomes invisible to {\it Swift} during 2010 
July 7 and July 28, the flux in X-rays decreases by more than
47 times within seven~days from the last detection of 2010 June
30. Although the present observations can give only upper 
limits to the decay timescale with $\Delta t_{\rm
dc} < 7\, {\rm days}$, and to the duration of being invisible
in X-rays with $21 \, {\rm days} \leq \Delta t_{\rm in} \leq 116 \,
{\rm days}$, due to gaps in the observed light curve, we still
have $\Delta t_{\rm dc} \ll \Delta t_{\rm in}$ and can give severe
constraints on the possible models for the origin of the
intermittence of the tidal flare. Here we explore in detail the
SMBHB model as suggested by LLC09, and give a discussion on
alternatives in Section~\ref{sec:alt}.

In the SMBHB model (LLC09), the orbits of the less-bound
  and unbound stellar plasma change due to the strong
perturbation by a massive object at a distance of several hundreds of
Schwarzschild radii. Then, a significant fraction of the stellar
plasma  cannot fall back to form an accretion disk and to fuel the
SMBH, leading to an intermittence of accretion onto the
SMBH, and thus of the tidal flare. Because the massive perturber
must be compact in order for it to survive the tidal force of the
central SMBH, it is most probably a second SMBH.
Before we numerically show how well the observations of SDSS
J1201+30 can be reproduced by the theoretical light curves
predicted by the SMBHB model in the Section~\ref{mod}, we first 
  briefly introduce and discuss the merits of the SMBHB model. 

After a star is tidally disrupted by a SMBH, the viscous
interaction among the stellar plasma is negligible and the plasma
fluid elements move ballistically with Keplerian orbits
before they fall back to the pericenter and circularize because
of strong shocks among the stellar tidal plasma \citep{eva89}. For the
ultra-bound fluid elements of Keplerian orbits with semi-major
axis, much less than that of the SMBHB (hierarchical triple systems),
the orbits of the fluid elements evolve secularly under the
perturbation of the secondary SMBH on the Kozai-Lidov time scale
\citep{koz62,lid62}, which is much larger than the Keplerian
  periods of the stellar fluid elements. The orbital angular momentum
change, within one Keplerian period of the fluid elements,
due to the torque caused by the quadrupole force of the secondary
is
\begin{equation}
  \delta{J_{\rm fl}} \sim {3\over 4} {G M_2 a_{\rm fl} \over a_{\rm
      b}^3} a_{\rm fl} 2\pi \left({a_{\rm fl}^3 \over G M_1}\right)^{1/2} ,
\end{equation}
where $a_{\rm fl}$ and $a_{\rm b}$ are, respectively, the semi-major
axis of the fluid elements and the SMBHB. Because the relative
change of the orbital angular momentum of the fluid elements
\begin{eqnarray}
  {\delta{J_{\rm fl}} \over J_{\rm fl}} & \sim & { 3\pi \over 2} {q (G M_1 a_{\rm
      fl})^{1/2} \over (G M_1 2 r_{\rm p})^{1/2}} \left({a_{\rm fl} \over
    a_{\rm b}}\right)^{3} \nonumber \\
      &\sim& 2.2\times 10^{-3} \beta^{1/2} q_{-1} \left({10 a_{\rm fl} \over
      a_{\rm b}}\right)^{7/2} \left({a_{\rm b} \over 10^{-3} \, {\rm
      pc}}\right)^{1/2} m_6^{-1/6} r_*^{-1/2} m_*^{1/6}
\label{eq:amratio}
\end{eqnarray}
is very small, the fluid elements must fall back to the pericenter and
accrete onto the SMBH as in a single SMBH system. In
Equation~(\ref{eq:amratio}),  $q = M_2/M_1 =0.1 q_{-1}$ is the ratio
of masses of the secondary and primary SMBHs.

However, when the fluid elements are less bound with semi-major
axes $a_{\rm fl} \ga a_{\rm b}/3$, the analytical and numerical
simulations show that the orbits of the fluid elements become chaotic
and change significantly on a Keplerian timescale \citep{mar01,liu09}
due to the nonlinear overlap of the multiple resonances \citep{mar07}.
From Equation~(\ref{eq:amratio}), we obtain the relative change
of the orbital angular momentum of the fluid elements with $a_{\rm fl}
\sim a_{\rm b}$ on Keplerian time
\begin{eqnarray}
  {\delta{J_{\rm fl}} \over J_{\rm fl}} &\sim& 7.1
  \beta^{1/2} q_{-1} \left({a_{\rm fl} \over a_{\rm
  b}}\right)^{7/2} \left({a_{\rm b} \over 10^{-3} \, {\rm
  pc}}\right)^{1/2} m_6^{-1/6} r_*^{-1/2} m_*^{1/6} ,
\label{eq:amcha}
\end{eqnarray}
and the fluid elements would move with chaotic orbits of pericenter
$r_{\rm p,fl} \sim 110 r_{\rm p}$, which is much larger than 
the pericenter $r_{\rm p}$ at disruption. Therefore, the fluid
elements move with chaotic orbits significantly different from one
another. They neither fall back to the pericenter at disruption,
nor cross each other to form shocks and accretion. The transition
radius, or corresponding Keplerian time $T_{\rm tr}$ of the fluid
elements from secular to chaotic orbits, depends critically on the
orbital parameters of the stars and also of the SMBHB
system, but $T_{\rm tr}$ can relate to the SMBHB orbit period, 
with $T_{\rm tr} = \eta T_{\rm b}$ (LLC09). The
parameter $\eta$ is a strong function of the orbital parameters of
the star, but is insensitive to the SMBH masses
(LLC09). Statistically, $\eta$ is in the range $0.15 \la \eta
\la 0.5$ with a mean value $\eta \sim 0.25$ for a SMBHB system
with eccentricity $e_{\rm b}=0$. Therefore, if $\eta$ is known,
one can determine the orbital parameters of the star
and SMBHB system. Taking into account that the first detection
of the tidal flare in SDSS J1201+30 is about $11.8\, {\rm days}$ after
the tidal disruption of the star and that the X-ray flux becomes
invisible at a time between 20 and 27 days after discovery, we
determine that
the observed truncation time is $31.8 \, {\rm days} \leq T_{\rm
  tr, obs} \leq 38.8 \, {\rm days}$ in the observer frame, or
$27.7 \, {\rm days} \leq T_{\rm tr} \leq 33.9 \, {\rm
  days}$ in the object rest frame.

When the stellar plasma stops falling back and fueling the
accretion disk of the outer radius $r_{\rm d} \approx r_{\rm c}$,
the accretion disk evolves dynamically on the viscous timescale at
$r \approx r_{\rm d}$
\begin{eqnarray}
  t_{\rm \nu, d} & = & {2\over 3} {r_{\rm d}^2 \over \nu(r_{\rm d})}
  \simeq {2\over 3} \alpha^{-1}
  \left(\frac{r_{\rm d}^3}{G\mbh}\right)^{1/2}
  \left(\frac{H}{r}\right)_{\rm r=r_d}^{-2}
  \nonumber\\
  &\simeq& 0.12~{\rm days} \times \alpha_{-1}^{-1} \left({\beta\over
  2}\right)^{-3/2} \left(\frac{H}{r}\right)_{\rm r=r_d}^{-2} r_*^{3/2}
  m_*^{-1/2} ,
\label{eq:tvisc}
\end{eqnarray}
where $\alpha = 0.1\alpha_{-1}$ is the disc viscosity coefficient and
$H$ is the scaleheight of the disk with $H\simeq r$ for the accretion
rate $\dot{m} \ga 0.3 \dot{M}_{\rm Edd}$
\citep{sha73,abr88,str09}. Equation~(\ref{eq:tvisc}) suggests 
that $t_{\rm \nu, d}$ is independent of the mass of the central
SMBH. For SDSS J1201+30, Equations~(\ref{eq:Mdot}) and (\ref{eq:tmin})
suggest that the accretion rate $\dot{M}$ is much larger than
$\dot{M}_{\rm Edd}$ when the source was absent from the {\it Swift}
observation on 2010 July 7, or 33.9 days after tidal
  disruption. Therefore, the
accretion disk in SDSS J1201+30 dynamically evolves on the viscous
timescale $t_{\rm \nu,d} \sim 0.12~{\rm days}$ for a typical viscous
parameter $\alpha = 0.1$. This is about 60 times less than the time
interval between the last detection on 2010 June 30 and the first
absence from the {\it Swift} observation on 2010 July 7 (S2012). 

For the dynamic evolution of the accretion disk without the fueling
of stellar plasma, here we construct a simple model to describe it,
by assuming that $\nu$ is only a function of radius with $\nu \propto
r^n$. For the physical model of an accretion disk in tidal
disruption as proposed by \citet{str09}, a kinematical viscosity
$\nu = \alpha c_s H \propto r^n$ with $n = 1/2$ is a good
approximation and adopted in the model. For an accretion disk
with $\nu \propto r^n$, the dynamic evolution can be
solved analytically and the surface density $\Sigma$ can be given by
making use of a Green's function $G(r,t)$
\begin{eqnarray}
  \Sigma(r,t) & = & \int_{r_{\rm in}}^{r_{\rm d}} G(r,R,t)
  \Sigma(R,t=t_{\rm tr}) dR ,
\label{eq:surf}
\end{eqnarray}
\citep{lyn74,tan11}, where $\Sigma(R,t=t_{tr})$ is the initial disk
surface density when the stellar plasma stops fueling the
accretion disk at $t=t_{\rm tr}$, $r_{\rm in}$ ($r_{\rm in} \simeq 3
r_{\rm g}$ for a Schwarzschild black hole) is the inner radius of
the accretion disk, and the Green's function $G(r,R,t)$ is
\begin{eqnarray}
G(r,R,t) & \approx & {2-n \over \Gamma(j+1)} \left({r\over r_{\rm
    d}}\right)^{-n} r_{\rm d}^{-5/2} R^{3/2} \tau^{-1-j}
    \exp\left[-{\left(r/r_{\rm d}\right)^{2-n} +\left(R/r_{\rm
    d}\right)^{2-n} \over \tau}\right]
\label{eq:green}
\end{eqnarray}
for $t-t_{\rm tr} \ga t_\nu(r_{\rm d})$. Here,
$j=(4-2n)^{-1}$, $\tau = 2(2-n)^2 (t-t_{\rm tr})/t_{\rm \nu,d}$,
and $\Gamma(j)$ is the Gamma function. At the moment, when the
stellar plasma stops fueling the accretion disk, the
surface density is approximately
\begin{equation}
  \Sigma(R,t=t_{\rm tr}) \approx \Sigma_{\rm d} \left(R/r_{\rm
    d}\right)^{-n} , \quad {\rm for} \quad r_{\rm in} \leq R \leq
    r_{\rm d} 
\label{eq:surf_in}
\end{equation}
where $\Sigma_{\rm d} \approx \dot{M}_0 / (3\pi \nu(r_{\rm d}))$ and
$\dot{M}_0$ is the accretion rate at time $t=t_{\rm
  tr}$. From Equations~(\ref{eq:surf}), (\ref{eq:green}), and
(\ref{eq:surf_in}), we have
\begin{eqnarray}
  \Sigma(r,t) & \approx & {2(2-n) \over (5-2n) \Gamma(j+1)}
  \Sigma_{\rm d} \left({r \over r_{\rm d}}\right)^{-n} \left[2(2-n)^2
  {(t-t_{\rm tr}) \over t_{\rm \nu,d}}\right]^{-j-1}
  \left[1-\left({r_{\rm in} \over r_{\rm
  d}}\right)^{{5\over2}-n}\right] ,
\end{eqnarray}
which gives an inward radial mass flow
\begin{eqnarray}
  \dot{M} = 6\pi r^{1/2} \frac{\partial}{\partial r}\left(\nu\Sigma r^{1/2}\right)
  \approx {2(2-n) \over (5-2n) \Gamma(j+1)} \dot{M}_0
  \left[1-\left({r_{\rm in} \over r_{\rm
  d}}\right)^{{5\over2}-n}\right] \left[2(2-n)^2
  {(t-t_{\rm tr}) \over t_{\rm \nu,d}}\right]^{-j-1}
\label{eq:accr1}
\end{eqnarray}
for $(t - t_{\rm tr}) \ga t_{\rm \nu, d}$. For a typical value
$n=1/2$, we have
\begin{eqnarray}
  \dot{M}(t-t_{\rm tr} \ga t_{\rm \nu,d}) \approx 0.113 \, \dot{M}_0
  \left[1-\left({r_{\rm in} \over r_{\rm
  d}}\right)^2\right] \left[{t-t_{\rm tr} \over t_{\rm
  \nu,d}}\right]^{-4/3} .
\label{eq:accr2}
\end{eqnarray}
Equation~(\ref{eq:accr2}) shows that the accretion rate in SDSS
J1201+30 decreases 50 times within about $3.7 t_{\rm \nu,d} \sim 0.4
- 1\, {\rm days}$ after the stellar plasma  stops fueling
the accretion disk. One has to notice that the accretion disk changes
the accretion mode to advection dominated accretion flows (ADAFs) for
$\dot{M} \la 10^{-2} \dot{M}_{\rm Edd}$ \citep{nar94} and our simple
model cannot be applied to the transition. Because an ADAF is
radiatively inefficient, it is expected that the object becomes even
dimmer.

The numerical simulations by LLC09 suggested that after a
time interval, the stellar plasma starts refueling the accretion
disk and the tidal disruption flare recurs. The exact time when the
tidal flare recurs depends on both the parameters of the
star and the SMBHB system, but it is about at time $T_{\rm r} =
\xi T_{\rm b}$ with $\xi$ of order unity. Given the gaps in the
light curve of \sdss,  the recurrence of the tidal flare can happen
between 2010 July 28 and
October 24, suggesting a recurrent time $T_{\rm r, obs}$
between $60 \, {\rm days}$ and $148 \, {\rm days}$ in the observer
frame or $T_{\rm r}$ between $52 \, {\rm days}$ and $129 \, {\rm
  days}$ in the object rest frame. Therefore, the orbital period,
$T_{\rm b}$ ($\sim T_{\rm r}$), of the
SMBHB in SDSS J1201+30 is on order of, or less than, a few hundred
days. The suggested orbital period of the SMBHB, together with $28\,
{\rm days} < T_{\rm tr} < 34 \, {\rm days}$, suggests $0.2 \la \eta
\la 0.6$, which is consistent with the prediction of the SMBHB model.
Whether and when the recurrent flare is interrupted again depends
not only on the orbital parameters of the star and SMBH
system, but also on the mass ratio and the total masses of the
SMBHs, which have to be determined numerically (LLC09).

%%---------------------------

\subsection{Numerical simulations and results}
\label{mod}

In Section~\ref{intr:an}, we showed that the observations of the tidal
flare in SDSS J1201+30 are consistent over all with the
predictions of the SMBHB model of LLC09.
In this section, we numerically calculate the model light curves for
the different parameters and quantitatively compare them one by
one with the light curve of SDSS J1201+30 in X-rays.
Given the sampling of the light curve of the tidal flare,
we do not use the least-square method to get the best fit of the model
predictions to the observations, but instead numerically explore
the model parameter space to obtain the range of the parameters with
which the SMBHB model can give theoretical light curves
reasonably reproducing the observations of SDSS J1201+30.

In the simulations, for simplicity we assume that the
disrupted star is a solar-type main-sequence star of
parameter $k=2.5$, and that the SMBHs is a Schwarzschild
black hole, which is consistent with the suggestions given by the
correlation of black hole spin and jet power \citep{nar12}. In this
section, we consider SMBHB systems of circular orbit with $e_{\rm b}
=0$, while the effects of the orbital eccentricity of the SMBHB
are investigated in Section~\ref{ecc}. The other parameters, both of
the disrupted star and the SMBHB system, are explored and
tested in a large range of values.

The method of the numerical simulations used in this work was
described in detail in LLC09. Here we give only a
  summary and interested readers are referred to that paper.
Because the interaction between the fluid elements is negligible
before they return to the pericenter of the disrupted star and
experience strong shocks because of the intersection and
  collision of the fluid elements with one another near pericenter,
we can simulate the orbital evolution of the fluid elements with
scattering experiments and calculate the fallback and
accretion rate of the stellar plasma onto the central SMBH, by
assuming that all the stellar plasma falling back to
a radius of around 2$r_{\rm p}$ or less is accreted. In each
scattering experiment, the mass of the star
is divided into fluid elements with specific energy $E_{\rm fl}$
and constant pericenter $r_{\rm p}$, by assuming a constant distribution
of mass in the specific energy in a range of $-\Delta E$ and $\Delta
E$, which is given by Equation~(\ref{eq:DEp}) with $k=2.5$. When the
fluid elements moving in the pseudo-Newtonian potential return and
pass by the central SMBH within a distance $2 r_{\rm p}$, the
calculation stops and
the fluid element and the time are recorded, respectively, as accreted
matter and accretion time. Otherwise, the calculation continues until
time $4 T_{\rm b}$ (the largest timescale in which we are
interested) or the fluid element escapes from the system. The
accretion rate of the stellar plasma is computed using the
recorded fluid element masses and accretion time.
We then compare the model accretion rate and the observations by adjusting
the free parameter $f_{\rm x}$ in Equation~(\ref{eq:fl_mat}). To
reproduce the observations, we require a model
light curve (1) to have an interruption between 2010 June 30 and 
July 6, one day before the first observation of absence on
  2010 July 7; (2) that a recurrent flare begins between 2010 July 28 
and October 24, and ends between 2010 December 13 and 2011 March 5;
and (3) to have a second interruption during the period around
2011 April 1.

Because the SMBH masses cannot be determined uniquely in the
SMBHB model by fitting the observations of the tidal flare,
they have to be estimated separately and taken as input values
in the simulations. Based on the relationship of black hole mass and bulge
luminosity of the host galaxy, S2012 estimated a black
hole mass in the range
$3\times 10^5 M_\sun < M_{\rm BH} < 2\times 10^7 M_\sun$ with a higher
value of $M_{\rm BH} \sim 10^7 M_\sun$ preferred. We have only done
scattering experiments for three typical values of the mass of
the primary SMBH, $\mbh = 10^6\msun$, $5\times10^6\msun$, and
$10^7\msun$. For a given mass of the primary black hole, the mass of
the secondary is calculated with the mass ratio $q$, which is
determined by reproducing the observed light curve. The results of the
numerical simulations show that the interruption time $T_{\rm tr}$,
the recurrent time $T_{\rm r}$, and the duration of the recurrent
flare strongly not only depend on the orbital parameters of the
disrupted star at disruption, including (1) the penetration
factor $\beta$,
(2) the inclination angle $\theta $ between the orbital planes of the
star and the SMBHB, and (3) the longitude of ascending node $\Omega$
and the argument of pericenter $\omega$ of the stellar orbits relative
to the orbit plane of the SMBHBs. They also depend on the
orbital period of the SMBHB, the SMBH masses, and the black hole
mass ratio $q$. To obtain the possible model light curves
consistent with the observations, in principle we have to do a large
amount of high precision simulations to cover all the
space of the seven model parameters, which would consume a large
amount of computation resources and is prohibitive for present
computing power. However, the discussions in
Section~\ref{intr:an} suggested that the orbital period of the
SMBHB is of
order hundred days, which could be taken as a good initial value
of the orbital period $T_{\rm b}$ and used to reduce the
computations. LLC09 suggested that, with a given orbital period of
the SMBHB, the
interruption time $T_{\rm tr}$ and the characters of recurrent
flares depend sensitively on the stellar orbital parameters
$\theta$, $\Omega$, and $\omega$, but less on the parameters of
the SMBHB system. Therefore, we started the simulations from
taking $T_{\rm b} \sim 135\, {\rm days}$ and $q\sim 0.1$ as initial
values and carried out the simulations for different values of
$\theta\in [0, \pi]$, $\Omega \in [0, 2\pi)$, and $\omega \in [0,
    2\pi)$ with a resolution of $0.1\pi$ in all the parameter 
space to obtain the typical sets of parameter values giving model
light curves closest to (but not necessary fully reproducing)
the observations. After a large number of simulations, we found that
the model light curves obtained with both $\theta \sim 0.3 \pi$ and
$\theta \sim$ $0.5 \pi$ can
give a reasonable fit to the observations, and the present
observations of SDSS J1201+30 cannot tell which one is
preferred. Those simulations with $\theta < 0.3 \pi$ tend
  to give a longer duration for the first part of the light curve,
  which is inconsistent with the observed truncated light curve around
2010 July 7. For the simulations with $\theta > 0.5 \pi$, most of
the results cannot give a proper recurrence as observed in
\sdss. Therefore, our results prefer a disrupted star with a prograde
orbit relative to the SMBHB for \sdss. However, our model cannot
entirely exclude a disrupted star with a retrograde orbit, because there
are a few results with $\theta \gtrsim 0.5 \pi$ still marginally
consistent with the observation. Further observations with higher time
resolution may help to solve the ambiguity. The simulation results
suggest that the required values of the parameters of the SMBHB are
nearly independent of the values of either $\theta \sim 0.3 \pi$ or
$\theta \sim 0.5 \pi$, although the solutions for $\theta = 0.5 \pi$
are fine tuning, in the sense that the solutions become significantly
different from the observations for a small deviation of
$\theta$ from $\theta = 0.5 \pi$. Because we are mostly interested in
the values of the parameters of the SMBHB system and the
large gaps in the light curve do not warrant any 
fine tuning solution, we adopt the solutions of $\theta \sim 0.3
\pi$. After $\theta \sim 0.3 \pi$ is chosen, we obtain the
corresponding values of the other orbital parameters of the star:
$\Omega = 0.2 \pi$ and $\omega = 1.5 \pi$.
Again, our test simulation results show that the values of the
model parameters of the SMBHB system required to reproduce the 
observed light curve are nearly independent of the values
of $\Omega$ and $\omega$. Therefore, we will give the numerical
simulation
results for $\theta=0.3\pi$, $\Omega = 0.2\pi$, and $\omega = 1.5
\pi$. With the given values of $\theta$, $\Omega$, and $\omega$,
we have carried out the numerical simulations iteratively,
starting from the initial values $T_{\rm b} = 135 \, {\rm days}$,
$q=0.1$ and $\beta = 2$, for the penetration factor $\beta$ between
$1 \leq \beta \leq 6$, $1 \leq \beta \leq 3$ and $1 \leq \beta
\leq 2.5$ and for the orbital period $T_{\rm b}$ between $130 \,
{\rm days} \leq T_{\rm b} \leq 160\, {\rm days}$, $135\, {\rm days}
\leq T_{\rm b} \leq 170\, {\rm days}$, $120\, {\rm days} \leq T_{\rm
b} \leq 200\, {\rm days}$, respectively, for $\mbh =10^6\msun$, $\mbh
=5\times 10^6\msun$ and $\mbh =10^7\msun$, and the mass ratio $q$
between $0.03 < q < 0.9$, in order to obtain all the model
light curves fully consistent with the observations, including 
the positive detections and upper limits in X-ray flux.

%%%
\begin{figure}
\begin{center}
\includegraphics[width=0.6\textwidth,angle=0.]{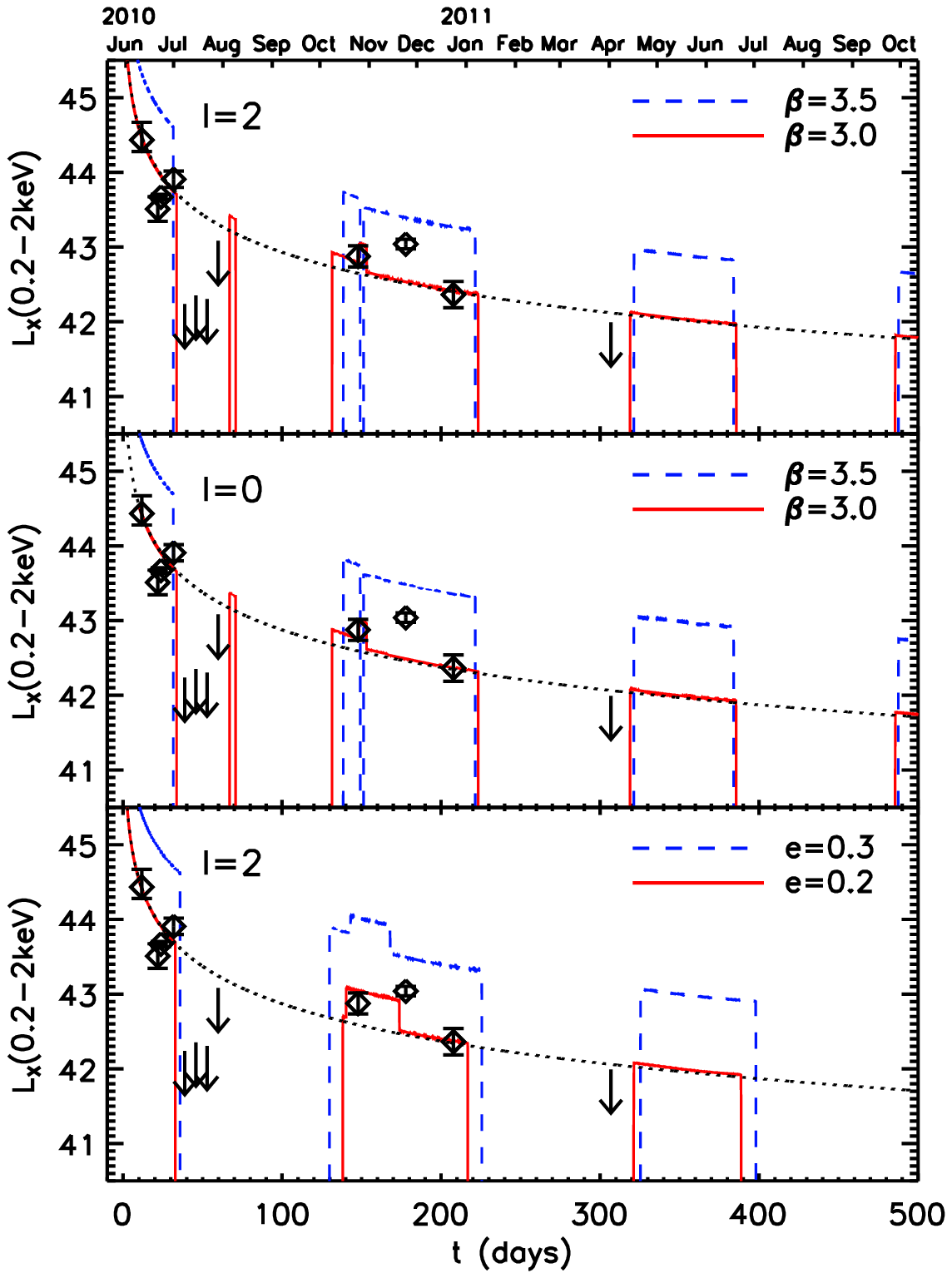}
\caption {
Simulated light curves in the observer frame
    for SDSS J1201+30 obtained with $M_{\rm BH} = 10^6 M_\sun$.
    Observational data from S2012 (open diamonds and arrows
    for upper limits) and a $t^{-5/3}$ power-law (dotted line) are
    overplotted. The top $x$-axis gives the dates of observation.
    In the simulations, $\theta = 0.3\pi$, $\Omega = 0.2\pi$, and
    $\omega = 1.5\pi$.
    Top panel: results for penetration factor
    $\beta=3.0$ (solid red line) and $\beta=3.5$ (dashed blue line,
    shifted upwards by 1 dex) with $l=2$. For both simulations,
    $e_{\rm b} = 0$, $T_{\rm b}=140 \, {\rm days}$, and $q=0.1$.
    Middle panel: same as top panel while $l=0$. The differences of
    the light curves between the top and middle panels are 
    $t_{\rm min}$ and $f_{\rm x}$ (see text for details). The first
    interruption in the light 
    curve for $\beta=3.0$ appears on 2010 July 2, about two days after
    the last detection and five days before the first upper limit.
    Bottom panel: results for the elliptical binary orbits with
    $e_{\rm b} =0.2$ (solid red line) and $e_{\rm b} =0.3$ (dashed
    blue line, shifted upwards by 1 dex), with $l=2$. For both
    simulations $\phi_{\rm b} = 1.7\pi$, $T_{\rm b}=150 \, {\rm
    days}$, $q=0.1$, and $\beta=2.0$. The first interruption in the
    light curve for $e_{\rm b} =0.2$ appears on 2010 July 4.}
\label{fig:M6} 
\end{center}
\end{figure}
%%%

We first carry out the numerical simulations for SMBHB systems
with a primary black hole of mass $M_{\rm BH} = 10^6 M_\sun$.
The top panel in Figure~\ref{fig:M6} gives two examples of the
model light curve for a penetration factor $\beta = 3.0$ (solid
red line) and $\beta=3.5$ (dashed blue line), with $l=2$, which
can fully reproduce the X-ray observations of the tidal flare 
in SDSS J1201+30. To obtain the model light curve in
Figure~\ref{fig:M6}, we have adopted a SMBHB with an orbital
eccentricity $e_{\rm b} = 0$ and a period $T_{\rm b} \simeq 140\, {\rm
days}$, and a black hole mass ratio $q \simeq 0.1$. To compare the
theoretical light curve and the observations in the X-rays, we
overplot the observational X-ray data and shift the theoretical light
curve up and down in order to obtain a good fit by eyes, which gives a
scaling factor in Equation~(\ref{eq:fl_mat}), $f_{\rm x} = 0.004$. For a 
typical transfer efficiency of matter to radiation $\epsilon = 0.1$,
this gives a total mass accreted $\Delta{M}_{\rm acc} \sim 0.02
\msun$, very small but consistent with the reported values for TDEs
in the literature \citep[e.g.,][]{kom02,li02,kom04,esq08,gez09,cap09,mak10,don14}. 
We recall that a number of unknowns determines the precise amount of
X-ray emission in the early evolution of TDEs and the unobserved EUV
component may dominate the bolometric luminosity, as discussed in 
Section~\ref{obs}. Therefore the value we report here just serves as
an order-of-magnitude estimate and lower limit. 
We notice that the orbital period of the SMBHB, $T_{\rm b} \simeq
140 \, {\rm days}$, obtained by detailed numerical simulations, is
consistent with the simple estimation given in
Section~\ref{intr:an}. The period $T_{\rm b} \simeq 140 \, {\rm
days}$ together with the black hole mass ratio $q=0.1$, suggests that the
SMBHB in SDSS J1201+30  has an orbital radius $a_{\rm b} \simeq 0.26
M_6^{1/3} \, {\rm mpc} \simeq 2.8 \times 10^3 M_6^{-2/3} r_{\rm g}$.
Figure~\ref{fig:M6}
shows that the recurrent flares in the SMBHB model light curve fit the
observations of SDSS J1201+30 in the X-rays better than a simple
$t^{-5/3}$-power law decay, except for the observation on 2010
November 23, which was strongly affected by radiation (S2012).

For comparison, we re-do the simulations in Figure~\ref{fig:M6}
with the same parameters, except $l=0$. The 
results have been plotted in the middle panel of
Figure~\ref{fig:M6}. Comparing to the top panel, there are only
small differences at the beginning of the light curve. Based on
Equation~(\ref{eq:DEb}), the energy dispersion for $l=0$ is smaller
than $l=2$ when $\beta > 1$, which results in a larger $t_{\rm
min}$. However, those changed $t_{\rm min}$ in Figure~\ref{fig:M6}
are still consistent with the observation. Besides, as implied
by Equations~(\ref{eq:Mdot}) and (\ref{eq:fl_mat}), since $\dot{M}(t)
\propto t_{\rm min}^{2/3}$, a larger $t_{\rm min}$ will lead to a
smaller $f_{\rm x}$. As a result, $f_{\rm x} = 0.0004$ for the
middle panel of Figure~\ref{fig:M6}, which is an order of magnitude
lower than for the model of the top panel. As shown in
the  middle panel, however, $\Delta E$ is independent of $\beta$ when
$l=0$, and the overall light curve is still impacted by $\beta$,
because the truncation and recurrence of the light curve in the SMBHB
system are sensitively depending on the size of the circularized
accretion disk, which should be changed for a different $\beta$ value.

From Figure~\ref{fig:M6}, the first interruption in the model light
curve for $\beta=3.0$ happens at about time $T_{\rm tr,obs} =
34$~days after the tidal disruption of the star, or on 2010 July 2,
which is about two days after the last detections on June 30 or five
days before the first upper limit of the {\it Swift} observation
  on 2010 July 7. Equation~(\ref{eq:accr2}) suggests a decay of
  the accretion rate by about a factor 600 within five days. This
  is consistent with the observation of a factor $>$~48
  within seven days. We note that for such a large decrease of the
  accretion rate, the simple dynamic model of the accretion disk may
not be applicable anymore. The theoretical light curve for $\beta =
3.0$ suggests that the stellar plasma stops returning and fueling
the accretion disk for a period $\Delta{T}_{\rm tr,obs} \approx 98\,
{\rm days}$, intermingling with a flicker of accretion for about
four days around 2010 August 7. The stellar plasma restarts fueling the
accretion disk at about 131~days after the tidal disruption or about
16~days before the detection with {\it Swift} on 2010 October 24. 
After about $92\, {\rm days}$, the recurrent tidal flare is
interrupted again, at about 16~days after the last detection with
{\it XMM-Newton} on 2010 December 23. This model predicts  that
the interrupted tidal flare will recur again later.

The results given in Figure~\ref{fig:M6} show that the observed
X-ray light curve of SDSS J1201+30 can be fully reproduced by the
theoretical light curve obtained with the typical model parameters
$\theta = 0.3\pi$, $\Omega = 0.2\pi$, $\omega = 1.5\pi$, $\beta =
3.0$, $M_{\rm BH} = 10^6 M_\sun$, $q= 0.1$, $e_{\rm b} = 0$, and
$T_{\rm b} = 140\, {\rm days}$. While our simulation results also
show that the observed light curve can be well reproduced by
the theoretical light curves obtained with model parameters in a
broader range of values, namely, $1.2 \la \beta \la 5.0$, $ 132\, {\rm
  days} \la T_{\rm b} \la 145\, {\rm days}$ and $0.05 \la q \la
0.18$. The upper panel of Figure~\ref{fig:M6} also gives the
theoretical light curve obtained with penetration factor $\beta =
3.5$ (dashed blue line). Values of the other model parameters
are the same as for $\beta = 3.0$. Figure~\ref{fig:M6} shows that a
small change of the penetration factor $\beta$ from $\beta = 3.0$ to
$\beta = 3.5$ would result in significant changes not only of the
time and duration of interruption, but also of the variability of
the recurrent flares. The simulation results show that the
variations of the penetration factor $\beta$ can significantly
change the timescale $t_{\rm min}$, interruption and recurrent time,
the duration of the recurrent flare, and the secondary recurrent
time. In fact, all the key signatures in the light curves for
SMBHBs strongly depend on complex combinations of the
parameters $\beta$, $T_{\rm b}$, and $q$. This implies that the
orbital parameters of the star and the SMBHB system, together
with black hole mass ratio (except for the mass of the primary
SMBH), can be determined observationally, when one can obtain
very densely sampled TDE light curves in the future. A deeper
exploration of the allowed values in parameter space is given in
Section~\ref{sec:para}.

Because the mass of the primary BH in \sdss is loosely
constrained by observations (approximately within the range
$3\times 10^5 M_\sun$ and $ 2\times 10^7 M_\sun$), we have also
carried out
the numerical simulations for SMBHB systems with black hole mass $\mbh
= 5\times10^6 M_\sun$ and $\mbh = 10^7 M_\sun$, and compare the
model light curves with the X-ray observations of \sdss. Again, here
we only consider SMBHB systems with circular orbit. The simulations
are carried out for model parameters in the ranges $0.03 \leq q \leq
0.9$,
$1 \leq \beta \leq 3$ and $1 \leq \beta \leq 2.5$, respectively, for
$\mbh = 5\times10^6 M_\sun$ and $\mbh = 10^7 M_\sun$. The simulation
results show that for a SMBHB system with $M_{\rm BH}=5\times
10^6\msun$ and $e_{\rm b} = 0$ a set of fine-tuned values of
model parameters
can give theoretical light curves consistent with the observations.
However, for a SMBHB system with $\mbh = 10^7 M_\sun$ and $e_{\rm b} =
0$, no theoretical light curve is found to be fully consistent with
the observations, in the sense that either the predicted first
interruption date is later than the first upper limit on July 7 or the
predicted recurrent flares do not cover all the observations from 2010
October 24 to December 23. As a result, if the orbit of the
SMBHB in \sdss is circular, the primary SMBH most probably
has a mass $M_{\rm BH} \la 5\times 10^6 M_\sun$. Otherwise, if the SMBH has
a mass $M_{\rm BH} \sim 10^7 M_\sun$ as preferred by the
observations (S2012), the SMBHB is hardly circular, however, 
because of the limited coverage of the parameter space the
present simulations cannot completely rule out the possibility that
fine-tuning the model parameters might give solutions to the observational
light curve. We will explore the effects of orbital eccentricity on the
  results in the next section.

\subsection{Orbital eccentricity of the SMBHB system}
\label{ecc}

The orbits of SMBHBs in galactic nuclei could be elliptical as
suggested by some models for the formation and evolution of SMBHBs in
galaxies \citep{arm05,ber06,pre11,iwa11,ses11,che11}. Because the
eccentricity of SMBHBs is of great
importance in modeling the templates of gravitational waveforms, we
now investigate if the observations of SDSS J1201+30 can give any
constraint on the orbital eccentricity. For an elliptical orbit, in
addition to the model parameters described in Section~\ref{mod}, we
now have two more parameters, the eccentricity $e_{\rm b}$ and the
initial phase $\phi_{\rm b}$ of the SMBHB orbit at the disruption of
the star.
In the simulations, $\phi_{\rm b}$ is set to be zero when the
two SMBHs are at closest distance. We have carried out the
scattering experiments by using the methods given in
Section~\ref{mod} and varying the eccentricity and initial
orbital phase in the ranges of $0 \leq e_{\rm b} \leq
0.9$ and $0 \leq \phi_{\rm b} < 2\pi$, respectively. The simulation
results show that both the time of interruption and
recurrence, and the durations of the recurrent flares of the
model light curves significantly change with $e_{\rm b}$ and
$\phi_{\rm b}$.  If the mass of the primary SMBH is $M_{\rm
BH} = 10^6 M_\sun$, the observations of SDSS J1201+30 can be well
reproduced with the model light curves obtained with the eccentricity
$0\leq e_{\rm b} \la 0.5$ and initial orbital phase $1.3\pi \la
\phi_{\rm b} \la 2\pi$. The bottom panel of Figure~\ref{fig:M6} gives
two examples of model light curves for \sdss computed with $e_{\rm b}
= 0.1$, $e_{\rm b} = 0.2$ and $l=2$. The first interruption in the
light curve for $e_{\rm b} = 0.2$ happens on 2010 July 4, about four
days after the last detection and three days before the first upper
limit on 2010 July 7. The recurrent fluxes in the model light curves
fit the observations of \sdss better than a simple power-law decay.
To obtain the light curves in the bottom panel of
Figure~\ref{fig:M6}, we have used the parameters $T_{\rm b} \simeq
150\, {\rm days}$ (or $a_{\rm b} \simeq 0.28 \, {\rm mpc}$),
$\phi_{\rm b} = 1.7\pi$ and $q=0.1$ for the SMBHB system, and
$\beta=2.0$, $\theta = 0.3\pi$, $\Omega = 0.2\pi$, $\omega = 1.5\pi$
for the star. For $e_{\rm b} = 0.2$, the model light
curves obtained with $0.05 \la q \la 0.15$, $1.2 \la \beta \la 4.7$
and $140\, {\rm days} \la T_{\rm b} \la 155\, {\rm days}$, can
fully reproduce the observations of \sdss. We notice that the values
of $\theta$, $\Omega$, and $\omega$ used in the simulations are the same as
for circular SMBHB systems. In fact, to reproduce the observed
light curve of \sdss, we take $\theta = 0.3\pi$, $\Omega =
0.2\pi$, and $\omega = 1.5\pi$ in all the simulations for different
eccentricities. However, we have to adopt different mass ratios
and values of the SMBHB orbital parameters, because the interruption
and recurrence in the model light curves depend on a complex
combination of the parameters of the SMBHBs. In conclusion, the
present observations of \sdss can constrain the eccentricity of the
SMBHB only in the range $0 \leq e_{\rm b} \la 0.5$, if the primary
SMBH has a mass $10^6 M_\sun$.

%----------
\begin{figure}
\begin{center}
\includegraphics[width=0.9\textwidth,angle=0.]{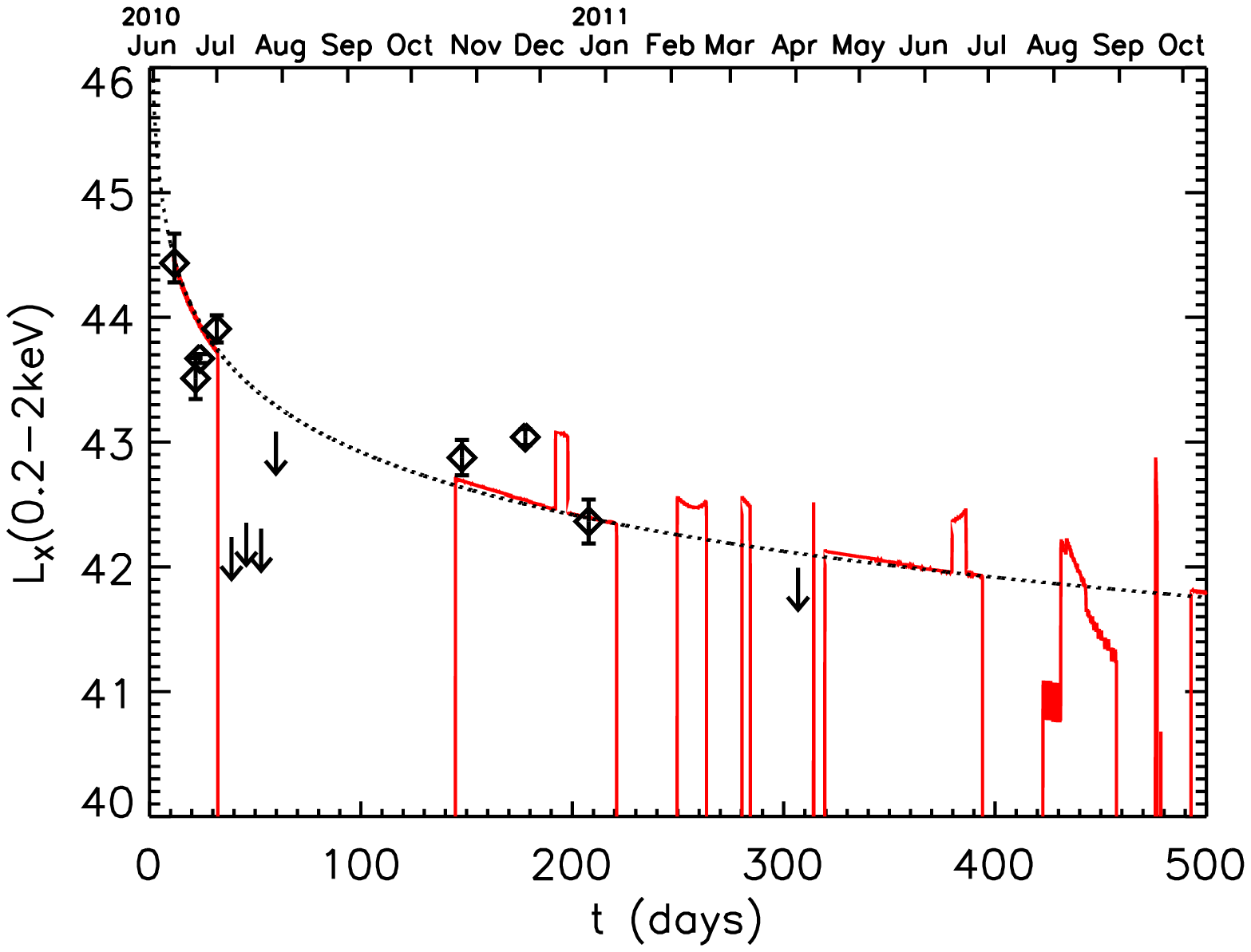}
\caption {Simulated light curve in the observer frame for
  \sdss obtained with $M_{\rm BH} = 10^7 M_\sun$ and eccentricity
  $e_{\rm b}= 0.3$. The top $x$-axis gives the observation
  dates. The model parameters include $l=2$, $\theta = 0.3\pi$,
  $\Omega = 0.2\pi$, $\omega = 1.5\pi$, $\beta=1.3$, $q=0.08$,
  $T_{\rm b}=150 \, {\rm days}$, and $\phi_{\rm b}=1.5\pi$. The
  observed data from S2012 (open diamonds and
  arrows for upper limits) and a
  $t^{-5/3}$ power-law (dotted line) are overplotted.
\label{fig:M7}}
\end{center}
\end{figure}
%-----------------

For $M_{\rm BH} = 5\times10^6 M_\sun$, the simulation results
show that the X-ray observations of \sdss can also be
reproduced with model light curves obtained with binary eccentricity
$0 \leq e_{\rm b} \la 0.5$. For a SMBHB system with orbital
eccentricities $e_{\rm b} \ga 0.5$, no model light curve is able to
reproduce the X-ray light curve of \sdss. For SMBHB systems with
an even higher SMBH mass of the primary, of $M_{\rm BH} \sim
10^7 M_\sun$, our results
show that we can obtain model light curves to reproduce the X-ray
observations of \sdss if the orbit of the SMBHB is elliptical
with eccentricity $0.1 \la e_{\rm b} \la 0.5$. If so, the
SMBHB system in \sdss would have an orbital period $140 \, {\rm
days} \la T_{\rm b} \la 160\, {\rm days}$ and a black hole mass ratio
$0.04 \la q \la 0.09$. The disrupted star would have a
penetration factor $1.3 \la \beta \la 1.6$ at disruption. The
orbital parameters then adopt the typical values: $\theta \simeq
0.3\pi$, $\Omega \simeq 0.2\pi$, $\omega \simeq 1.5\pi$, and
$\phi_{\rm b} \simeq 1.5\pi$. Figure~\ref{fig:M7} gives the simulated
light curves for \sdss by a SMBHB model with $M_{\rm BH} = 10^7 M_\sun$
and $e_{\rm b}= 0.3$. To obtain the light curve, we have used the
scaling factor $f_{\rm x} = 0.0004$, penetration factor $\beta = 1.3$,
black hole mass ratio $q=0.08$, and SMBHB orbital period $T_{\rm b} =
150\, {\rm days}$ or semi-major axis $a_{\rm b} \simeq 0.59 \, {\rm
  mpc} \simeq 620 r_{\rm g}$. Our results suggest that if the mass of
the SMBH in \sdss is indeed about $10^7 M_\sun$, as preferred by the
observations (S2012), the orbit of the SMBHB system should be
elliptical with moderate eccentricity $e_{\rm b} \sim 0.3$.

%%%%%%%
\section{Discussion}
\label{dis}

\subsection{SMBHB model and alternatives}
\label{sec:alt}

Several candidate small-separation SMBHBs have emerged in recent
years. They are not spatially resolved, but the SMBHBs' presence has
been indirectly inferred from semi-periodicities in light curves or
structures in radio jets (see Section~\ref{introduction} for the
observational evidence for SMBHBs; see also \citet{kom06} for a
complete review).

We have shown that the presence of a SMBHB can naturally explain
the characteristics of the light curve of the TDE in SDSS J1201+30,
based on models of LLC09. In fact, other possible
explanations that might at first glance come to mind, do not
well reproduce the features of the light curve, or the overall
multi-wavelength properties of SDSSJ1201+30.
We comment on each of them in turn.

\subsubsection{Jetted TDEs, and a comparison with SwiftJ1644+57}

Rapid, short-timescale variability has been detected in the light curve
of the TDE candidates Swift J1644+57 \citep[e.g.,][]{bur11,blo11} and
Swift J2058+0516 \citep{cen12}. However, unlike SDSS J1201+30, which has
no detectable radio emission, both Swift J1644+57 and Swift J2058+0516
are accompanied by strong radio emission, and their isotropic X-ray
luminosity, up to a few times $10^{47-48} {\rm erg\; s^{-1}}$, is well
above 
the Eddington limit. Therefore, beaming has been suggested to explain
these events \citep[e.g.,][]{bur11,blo11,lev11,zau11}. In particular,
it has been suggested that the rapid dips of the X-ray light curve of
Swift J1644+57 are linked to the presence of a jet, and are due to jet
precession/nutation \citep[e.g.,][]{sax12b}, or wobbling
\citep{tch13}. Since no radio jet was detected in SDSS J1201+30, it is
highly unlikely that similar scenarios are at work in this source.

\subsubsection{Temporary absorption due to blobs in the
  accretion disk, or stellar streams, or an expanding disk wind}

If an optically thick blob in the accretion disk extends vertically
above a slim disk and crosses the line-of-sight (l.o.s; the
disk is always optically thick in the
radial direction for the slim or standard thin disk in which we
  are interested), the absorption by the blob may lead to a sharp drop
in the flux. Because the largest Keplerian period at $r_{\rm d}$
is about 7.8 hr and independent of the black hole mass, the blob
should be extended to be an annulus in the phi-direction, if the entire
duration of the interruption period of at least a few days is due to the
eclipse. Such an annulus should be also extended in the radial
direction to the entire disk, in order for the eclipse time to last
about 21 days, at least 10--100 times the disk viscous timescale 
(cf. Equation~(\ref{eq:tvisc})). Therefore, in this scenario, the disk
must be a spherical accretion flow instead of a slim accretion disk. 
This is very unlikely because of the conservation of angular momentum.  

A blob with radius $r_{\rm b} \geq r_{\rm d}$ in the stellar
streams
may completely shield the accretion disk if it crosses the l.o.s. by
chance with a probability $P\sim 6 \times 10^{-4} \beta^{-3/2}
M_6^{-1/2} m_*^{1/2}$. If the blob is  a fraction $\lambda$ of the star
and not bound by self-gravity, it expands following the ejection of
unbound plasma from the system with volume $V(t) \approx 4\pi r_b^3
/3 \approx \lambda^3 R_*^3 \zeta^3$, where $\zeta = (t-t_{\rm D})/
t_{\rm e}$ is the expansion parameter and $t_{\rm e} \approx 1120 \,
{\rm s}\, \beta^{-4/3} m_*^{-1/2} r_*^{3/2}$ \citep{kas10}. At the
last detection of 2010 June 30, $\Delta t = t- t_{\rm D} \simeq 27.7
\, {\rm days}$ in the object rest frame suggests an expansion
parameter $\zeta \approx 2137$, or a blob with expanded size $r_{\rm
  b} \approx 6.6 \beta \lambda M_6^{-1/3} m_*^{1/3} r_{\rm d}$. For a
blob with $\lambda \sim 1$, the blob is large enough to fully shield
the tidal accretion disk and also still optically thick because of
Thomson scattering in ionized hydrogen for an expansion parameter
$\zeta \simeq 2000$ \citep{kas10}, although the probability of 
totally shielding the accretion disk is only
$P\sim 8\times 10^{-5} \lambda^2 \beta M_6^{-1/2} m_*^{1/2}$. However,
the timescale for the blob to cross and totally shield the
accretion disk is $\Delta t_{\rm ec} \sim 11\, {\rm days}
\beta^{1/2} M_6^{-1/6} m_*^{1/6} \left({\Delta t \over 27.7 \, {\rm
    days}}\right)$ in the object rest frame or 13~days in the observer
frame, which is about twice the observed upper limit to the
timescale of absence, $\leq 7\, {\rm days}$. Recent numerical
simulations suggested that for $\beta <3$ the unbound stellar
debris from the TDE remains self-gravitating and recollapses into thin
streams \citep{gui13}. Therefore, the expansion parameter $\zeta$ is $
< 1$ and any blob in the stream is unable to fully shield the tidal
accretion disk to form the extremely large dip in the light curve.

If an optically thick inhomogeneous clumpy structure in an
expanding, optically thin accretion-disk wind could form due to some
unknown reason after the accretion has passed the peak accretion
rate for some time, it may completely absorb the disk emission and form the
dips. However, to fully cover the accretion disk, the clumpy structure
must extend to as large as the accretion disk or the clumpy structure
in the disk wind is global. Because the clump is optically thick and
would radiate at the order of the Eddington luminosity with
a temperature about $10^5$~K before it becomes optically thin
\citep{str09}, it would emit a peak radiation at the UV and soft-X-ray 
bands and should have been detected with {\it XMM}.

\subsubsection{Transient eclipses due to stars or an optically-thick dense
    molecular cloud infalling toward the central SMBHs}

A giant star or a dense optically thick gas cloud infalling
toward the central
SMBHs with extent along the orbital plane $\Delta R \geq (21/7) 2
r_{\rm d} \approx 6 r_{\rm d}$ may transit and
completely shield the accretion disk to form a full
eclipse within seven days of 21 day duration, as was suggested for
some AGNs \citep[e.g.,][]{mck98,gil12,bek13}. For a giant star with
radius $R_* \simeq 100 R_\sun$, it requires a
SMBH mass $M_{\rm BH} \leq 4.6\times 10^3 M_\sun (\beta/2)^3 r_*^{-3}
m_*$ in order for the giant star to completely shield
the accretion disk. For a falling dense gas cloud to cross the tidal
accretion disk and form a total eclipse within $7/(1+z)$~days,
the dense gas
cloud should be at a distance from the central SMBH $r_{\rm gas} \la
1.6 \times 10^5 r_{\rm g} \beta^2 ({\Delta t \over 6\, {\rm days}})^2
M_6^{-2/3} r_*^{-2} m_*^{2/3}$ and have an extension along the
orbital plane $\Delta R_\| > 6 r_{\rm d}$ and vertical to the orbital
plane $\Delta R_\bot > 2 r_{\rm d}$. The dense gas cloud would
subtend a solid angle toward the SMBH $\Delta \Omega \ga \Delta R_\|
\Delta R_\bot /
r_{\rm gas}^2 \approx 1.1 \times 10^{-6} \beta^{-6} r_*^6 m_*^{-2}
({\Delta t \over 6\, {\rm days}})^{-4}$, implying a probability
$P \ga 9\times 10^{-8}$ for such a dense gas cloud to cross the
l.o.s. by chance. The probability can increase significantly if the
dense cloud has much larger size or the galactic nucleus is full of
small dense clouds similar to those in the broad-line
regions of AGNs. In both cases, the gas clouds should be ionized
by the tidal flare and emit strong broad emission
lines. However, these have not been detected in the optical spectra
taken at 12 days and 11 months after the discovery of the TDE
(S2012).

\subsubsection{Blobby accretion near last stable orbit or other
    AGN-like variability}

In rare cases, AGN light curves do show occasional rapid drops or
flares by factors of a few or more. Therefore, whatever causes
short-time variability in AGN accretion disks could
potentially also be at work in any TDE source, like in SDSS
J1201+30. However, it is clear that accretion disks in AGNs and tidal
disruption events are significantly different: the accretion disk in
AGNs can extend outward to several thousands of Schwarzschild radii
or more, but the size of the accretion disk in TDEs 
can only be as large as a few dozens of Schwarzschild radii (i.e., 
hundreds of times smaller). Therefore, the characteristic timescales of
the sources driving the variations in AGNs
\citep{kel09,kel11,tan10} range from a few hours up to a 
few hundreds of years for AGNs, but are always less than the
  viscous timescale $2.8\, {\rm hr}\, \alpha_{-1}^{-1}
(\beta/2)^{-3/2}$ in TDEs 
\citep[e.g.,][]{dec12}. The observational timescale of order of
days for \sdss is much larger than the characteristic timescales
of the driving sources in TDEs, which suggests stochastic
variations driven by blobs or white noises superimposing on the mean
luminosity as in AGNs \citep{kel09,kel11,tan10}, or in the tidal
disruption event {\it Swift 1644+57}
\citep{dec12}. If the fractional amplitude of the driving noises is
consistent with the observations of SDSS J1201+30 from 2010 June 10 to
June 28 and between 2010 October 24 and December 23, the probability
that one can detect the consecutive three observations at the extreme
quiescent state of the flare from 2010 July 7 to July 21
is $P\sim p_q^3 \la 10^{-15}$, where $p_q < 10^{-5}$ is the
probability to observe a variability with larger than $5 \sigma$
standard errors. If such large random variability with a
timescale less than a few thousand seconds did exist in the light
curve of SDSS J1201+30, it must have already been detected in the
{\it XMM-Newton} pointed observations with the exposure time ranging
from 10 ks to 30 ks on 2010 June 22, November 23, and December 23,
which, however, was not (S2012). 

\subsubsection{Lense-Thirring precession of an accretion disk
  misaligned with a spinning SMBH}

If the central SMBH is spinning and the accretion disk is misaligned
with the spin axis, the disk may precess about the total angular
momentum of the black hole accretion disk system as a solid body due
to the Lense-Thirring torque \citep{fra07}. The precession of
the accretion disk would lead to quasi periodic oscillations 
in TDE light curves \citep{dex11,sto12b,she13a} and to a
reduction of the observed flux up to about 50 times if the disk
precesses from a face-on into edge-on phases \citep{ulm99}. For an
accretion disk with surface density given by the
Equation~(\ref{eq:surf_in}), the precession period is
\begin{equation}
T_{\rm prec} \simeq {8\pi G M_{\rm BH} (1+2n) \over a c^3 (5-2n)}
{\left(r_{\rm d} /r_{\rm g}\right)^{5/2-n} \left(r_{\rm in} /r_{\rm
    g}\right)^{1/2+n} \left(1-(r_{\rm in} /r_{\rm d})^{5/2-n}\right)
  \over  1-(r_{\rm in} /r_{\rm d})^{1/2+n}}
\label{eq:pre}
\end{equation}
\citep{fra07,sto12b,she13a}, where $a$ is the dimensionless black hole
 spin parameter with $0 \leq a < 1$.
For a black hole mass $M_{\rm BH} = 10^7 M_\sun$ and typical disk
parameters $n=1/2$, $r_{\rm in} \simeq 3 r_{\rm g}$, and $r_{\rm d}
\simeq 2 r_{\rm t} \simeq 10 r_{\rm g}$, the typical disk precession period
is $T_{\rm prec} \approx 2.84 \, {\rm days}\, a^{-1} M_7$, or in the
observer frame, $T_{\rm prec, obs}= (1+z) T_{\rm prec} \approx 3.25 \,
{\rm days}\, a^{-1} M_7 \ga  3.25 \, {\rm days} \, M_7$. While for a 
black hole mass $M_{\rm BH} = 10^6 M_\sun$ and typical disk
parameters $n=1/2$, $r_{\rm in} \simeq 3 r_{\rm g}$, and $r_{\rm d}
\simeq 2 r_{\rm t} \simeq 47 r_{\rm g}$, the typical precession period
is $T_{\rm prec} \approx 5.2 \, {\rm days}\, a^{-1} M_6$, or in the
observer frame, $T_{\rm prec, obs}= (1+z) T_{\rm prec} \approx 5.9 \,
{\rm days}\, a^{-1} M_6$. If the large drop in the light curve of
\sdss is due to the disk Lense-Thirring precession with period
$T_{\rm prec, obs} \ga 3.25 \, {\rm days} \, M_7$, the observations
with regular time intervals of $10\, {\rm days}$ between 2010 June 10 
and June 30, and $7\, {\rm days}$ between 2010 July 7 and
July 28 of S2012 suggest that the face- and edge-on
phases should be, respectively, longer than 20 days and 21
 days. Therefore, the observations obtained between 2010 June 10 and
 July 28 imply that the precession period should have $T_{\rm
 prec, obs} \geq 48 \, {\rm days}$. However, the observations of the TDE
 between 2010 October 24 and December 23 suggest a face-on phase longer
 than 60~days and thus a precession period $T_{\rm prec, obs} \ga 81
 \, {\rm days}$. The observation of 2010 December 23 suggests $T_{\rm
 prec, obs} \geq 84.5 \, {\rm days}$, while the observation of 2010
 October 24 indicates $T_{\rm prec, obs} < 88 \, {\rm days}$. In
 conclusion, if the large drop in the flux was due to the disk
Lense-Thirring precession, the precession period would be $ 84.5 \,
{\rm days} < T_{\rm prec, obs} < 88 \, {\rm days}$. 
The drop in flux by a factor 50 within seven days, or $\la 8\%
$ of the precession period, is too steep to be consistent with the
predictions of Lense-Thirring precession \citep{dex11,sto12b,she13a}.

\subsection{Orbital parameters of the SMBHB system}
\label{sec:para}

All of our results were obtained with $\Delta E$ given by
Equation~(\ref{eq:DEp}) with $l=2$. However, some numerical simulations
\citep[e.g.,][]{sto12,gui12} suggest a weaker dependence of $\Delta
E$ on $\beta$. To investigate the dependence of our results on the
different relationships of $\Delta E$ and $\beta$, we have carried out
test numerical scattering experiments with $l=0$. 

The test numerical simulations are only carried out for $M_{\rm BH} =
10^6 M_\sun$, because the present results obtained with $\Delta E
\propto \beta^2$ show a large variation of $\beta$ with $1.2 \la \beta
\la 5$ 
for $M_{\rm BH} = 10^6 M_\sun$ but nearly a constant $\beta$ with $1.3
\la \beta \la 1.6$ for $M_{\rm BH} = 10^7 M_\sun$. 
Similar to Figure~\ref{fig:M6}, our results show that the
simulated light curves for \sdss 
and the obtained model parameters of the star and SMBHB systems
are nearly the same as those obtained with $l=2$, 
which can be understood as follow: different relationships of $\Delta
E$ and $\beta$ significantly change only $t_{\rm min}$, which is
degenerate with the scaling free parameter $f_{\rm x}$ as discussed in
Section~\ref{obs}. Meanwhile, although $\Delta E$ is independent of 
$\beta$ for $l=0$, the truncation and recurrence of the light curve
sensitively depends on $\beta$. Therefore, we will not discuss the
results obtained with $l=0$ any further.

\begin{figure}
\begin{center}
\includegraphics[width=0.8\textwidth,angle=0.]{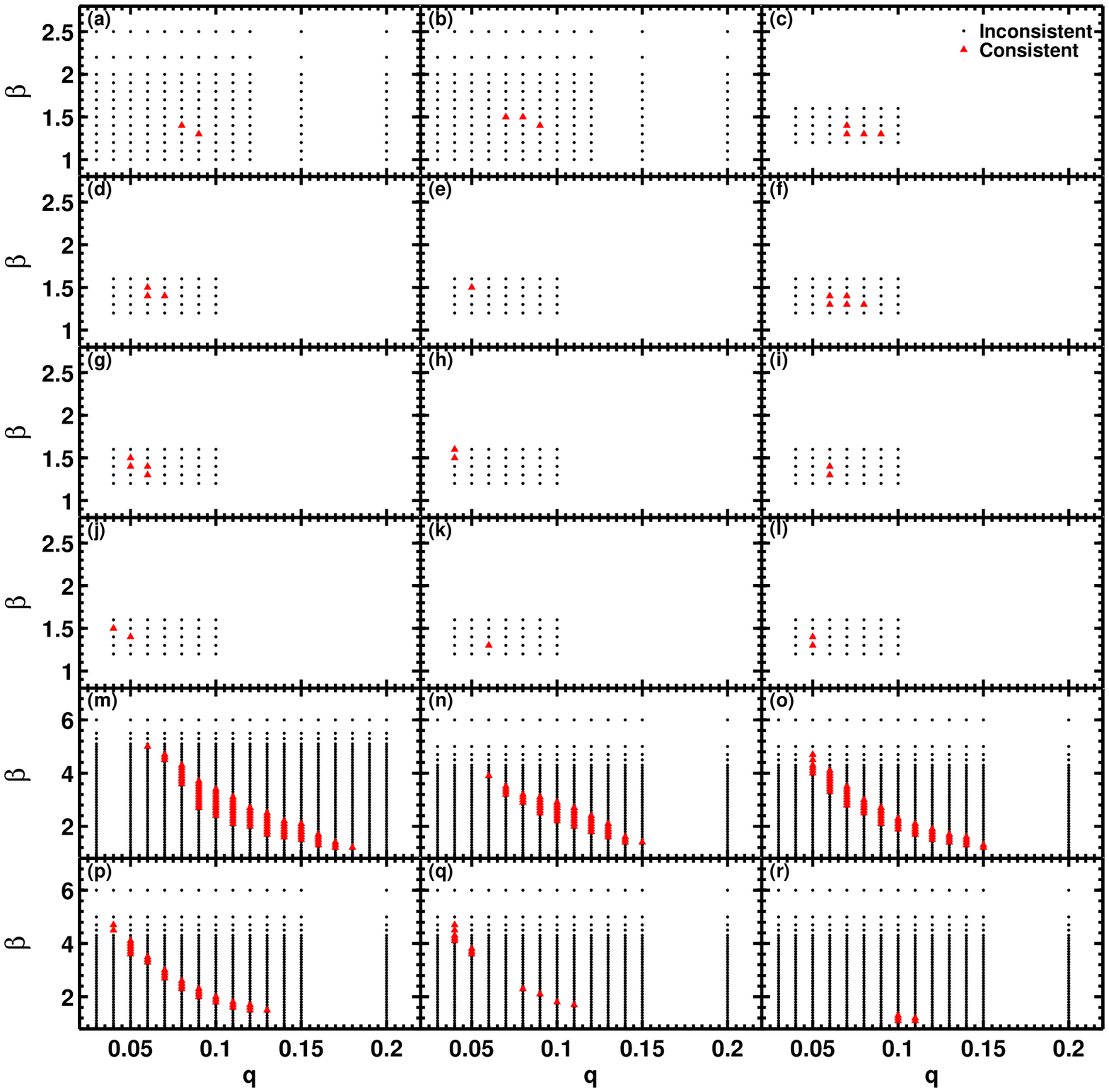}
\caption {Simulation results for the TDE of \sdss in the
    ($e_{\rm b}$, $q$, $\beta$) space for which the model light
    curves have been simulated. Numerical simulations have been done
    in the ($e_{\rm b}$, $q$, and $\beta$) space with $0 \leq
    e_{\rm b} \leq 0.9$, $0.03 \leq q \leq 0.9$, and $1 \leq
    \beta \leq 6$ for $\mbh=10^6\msun$ or $1 \leq \beta \leq 2.5$ for
    $\mbh=10^7\msun$. For ellipticity, those values, for which
    no model light curve is consistent with the observations of the
    TDE, are not shown here. In the computed domain of the
    parameter spaces, the fraction marked with the red triangle can
    give model light curves consistent with the observations of the
    TDE, while the other (marked with black dots) cannot.
    Panels (a)-(l) are results for $\mbh=10^7\msun$ with
    $\phi_{\rm b}=1.5 \pi$:  panels (a)-(b) for $e_{\rm b} =
    0.1$ and 0.2 with $T_{\rm b} = 140 \, {\rm days}$;
    panels (c)-(e) for $e_{\rm b} = 0.2$, 0.3, and 0.4 with
    $T_{\rm b} = 145 \, {\rm days}$; panels (f)-(h) for $e_{\rm
    b} = 0.3$, 0.4, and 0.5 with $T_{\rm b} = 150 \, {\rm days}$;
    panels (i)-(j) for $e_{\rm b} = 0.4$ and 0.5 with $T_{\rm b} = 155 \,
    {\rm days}$; panels (k)-(l) for $e_{\rm b} = 0.4$ and 0.5
    with $T_{\rm b} = 160 \, {\rm days}$.
    Panels (m)-(r) are for $\mbh=10^6\msun$ and
    $e_{\rm b} =0, 0.1,0.2,0.3,0.4$, and $0.5$, respectively. In the
    simulations, $T_{\rm b} = 140 \, {\rm days}$ for $e_{\rm b} =0$
    or $T_{\rm b} = 150\, {\rm days}$ and $\phi_{\rm b} = 1.7\pi$
    for $e_{\rm b} \ge 0.1$ are adopted.
\label{fig:all9}}
\end{center}
\end{figure}

We have shown that the peculiar characteristics of the
light curves of the TDE in \sdss can be naturally explained by the
SMBHB model given by LLC09. In the SMBHB model, the orbital
parameters of the disrupted star approaching with negligible
initial total energy can be determined to be $\theta
\sim 0.3\pi$, $\Omega \sim 0.2 \pi$, and $\omega \sim 1.5\pi$
independent of the detailed choices of the model parameters. By fully
reproducing the X-ray light curve of the TDE with the model light
curves of a SMBHB system, we have obtained that the orbit of the SMBHB
in \sdss should be elliptical with moderate eccentricity $e_{\rm _b}
\simeq 0.3$ (with $0.1 \la e_{\rm b} \la 0.5$) and typical orbital
period $T_{\rm b} \simeq 150\, {\rm days}$ (with $140 \, {\rm days} \la
T_{\rm b} \la 160 \, {\rm days}$) for a primary SMBH of mass $M_{\rm
BH} = 10^7 M_\sun$. A SMBH with that value is preferred by observations
(S2012). The orbital period of the SMBHB suggests a
semi-major axis $a_{\rm b} \simeq
0.59 \, {\rm mpc} \simeq  620 r_{\rm g}$. The black hole mass ratio,
the penetration factor of that star, and the initial orbital phase at
disruption have typical values $q\simeq 0.08$, $\beta\simeq 1.3$, and
$\phi_{\rm b} \simeq 1.5\pi$, respectively. Because the parameters
$q$, $e_{\rm b}$, and $\beta$ are three important model quantities and
the simulation results sensitively depend on their values, we have
carried out a large amount of simulations in three-dimensional
space. The results are given in the panels (a)-(l) of
Figure~\ref{fig:all9}. In the simulations, we
have adopted $\phi_{\rm b} = 1.5\pi$ for simplicity. The
simulation results show that the model solutions to the X-ray
light curve of the TDE consist only of a small domain in the
four-dimensional space ($\beta$, $q$, $e_{\rm b}$, $T_{\rm b}$) with
$T_{\rm b} \sim 150\, {\rm days}$. In panels (a)-(b) of
Figure~\ref{fig:all9}, we show the results obtained with
$T_{\rm b} = 140\, {\rm days}$ only for $e_{\rm b} = 0.1$ and 0.2,
and $0.03 \leq q \leq 0.2$, although we have done simulations for $0.1
\leq e_{\rm b} \leq 0.9$ and $0.03 \leq q \leq 0.9$, because no
model light curve obtained with parameters outside the ranges
can reproduce the X-ray light curve of the TDE. In panels (c)-(l),
we show the simulation results in the three-dimensional
space ($\beta$, $q$, $e_{\rm b}$) for $T_{\rm b}=145, 150, 155, 160\,
{\rm days}$. Because the model solutions are insensitive to the small
change of the orbital period, we have only done the simulations for
$0.1 \leq e_{\rm b} \leq 0.6$, $1.2 \leq \beta \leq 1.6$, and $0.03
\leq q \leq 0.09$.

%----------------------
%----------------

However, the mass of the central black hole can be smaller. If
the SMBH has a mass $M_{\rm BH} = 10^6 M_\sun$, we have shown that the
X-ray light curve of the TDE can also be reproduced with the model light
curves obtained with SMBHB systems of orbits either circular $e_{\rm
b} = 0$ or elliptical $e_{\rm b} \la 0.5$. The fitted orbital period
of the SMBHB depends weakly on the orbital eccentricities and has
a typical value of $T_{\rm b} \simeq 140 \, {\rm days}$ (or $a_{\rm
b} \simeq 0.26\, {\rm mpc}$) with $132\, {\rm days} \la T_{\rm b} \la
145\, {\rm days}$ if $e_{\rm b} = 0$ or
$T_{\rm b} \simeq 150 \, {\rm days}$ (or $a_{\rm b} \simeq 0.28\, {\rm
  mpc}$) with $142 \, {\rm days} \la
T_{\rm b} \la 156 \, {\rm days}$ if $e_{\rm b} = 0.2$. The fitted mass
ratio of the SMBHB system in \sdss has a typical value $q \sim 0.1$
for any possible orbit with $e_{\rm b} \la 0.5$, but distributes in
a range depending sensitively on the values of eccentricity and
penetration factor. The panels from (m) to (r) of
Figure~\ref{fig:all9} show the complex dependence among $q$, $e_{\rm
  b}$, and $\beta$ in the parameter space, which are adopted from our
simulation results for $0 \leq e_{\rm b} \leq 0.9$ and $0.03 \leq q \leq
0.9$. In the simulations, we have taken $T_{\rm b} = 140\, {\rm days}$
for $e_{\rm b} = 0$ and $T_{\rm b} = 150\, {\rm days}$ and $\phi_{\rm
  b} = 1.7 \pi$ for $e_{\rm b} > 0$ for simplicity. Different
orbital periods are adopted here for different types of orbits in
order to give the largest domain of model solutions in the parameter
space. Again, the results depend only weakly on the orbital
periods, in the sense that the variation of the orbital period in a
certain range only changes the boundaries, but not the shape of the
domain of the model solutions in the parameter space.

The model solutions of the observed X-ray light curve are
 obtained with the assumption that the disrupted star is a
solar-type main-sequence star. The disrupted star may be
a different type of star, with different internal structures, masses,
and/or radii. The tidal disruption of stars with different internal
structures would lead to different light curves with
different peak luminosities, peak time $t_{\rm peak}$,
and power-law indices other than $-5/3$ \citep{lod09,gui12}.
Because the key features of the interruptions and recurrences in
the model light curves do not change with the structure of the star,
our conclusions regarding the model solutions and the parameters are
robust. For a certain orbital
pericenter, Equation~(\ref{eq:DEp}) suggests the maximum specific
energy $\Delta E$ changes with the radius of the star. The differences
in $\Delta E$ do not change our conclusions as discussed before.
However, our results do sensitively depend on the orbital pericenter
$r_{\rm p}$ or $\beta$ for a certain tidal radius $r_{\rm t}$. If we
have the knowledge of the star, with the condition $r_{\rm p} \leq
r_{\rm t}$ or from Equation~(\ref{eq:beta}) we may give some more
strict constraints on $r_{\rm p}$ (or $\beta$) and thus the mass
ratio $q$ (and probably $e_{\rm b}$), based on Figure~\ref{fig:all9}.
Notice that in Figure~\ref{fig:all9} $\beta = \beta_\sun \equiv r_{\rm
t, \sun} / r_{\rm p}$ with $r_{\rm t, \sun}$ the tidal radius of a
 solar-type star.

We have carried out numerical simulations to scan a significant
fraction of the large model parameter space, but it is still
prohibitive for the present computing powers to cover the whole
space of the parameters ($\theta$, $\Omega$, $\omega$, $\beta$,
$T_{\rm b}$, $q$, $e_{\rm b}$, and $M_{\rm BH}$). From the 
first principles, the potential sphere of the restricted three-body
systems consisting of the SMBHB and the stellar plasma elements
is split into inner and outer regions by the SMBHB orbit. As
argued for Newtonian potential analytically by \citet{mar07} and
  numerically by LLC09, the shell of the potential sphere centered on
  the SMBHB orbit is chaotic because of the nonlinear overlap of the
  multiple resonances due to the strong perturbations by the secondary
  SMBH. The fluid elements with orbits inside the chaotic shell
  regions are resonantly scattered off and thus the inner and outer
  boundaries of the chaotic shell regions determine, respectively, the
  interruption and recurrence time of the TDE light curve. For a
  circular SMBHB orbit, the orbit of the SMBHB binary roughly
  determines the centers of the chaotic regions, suggesting that the
  Keplerian orbital period of the SMBHB is roughly between the time of
  the interruption and recurrence. The BH mass ratio determines the
  strength of the resonant perturbation, the extent of the
  nonlinear overlap of multiple resonances, and how effectively the
  fluid elements with orbits inside or crossing the chaotic shell
  regions are scattered off on the timescale when they move inside the
  chaotic shell, deciding the duration and shapes (e.g., smooth or
  flickering) of the interruptions and the recurrent light curves.
  The orbital parameters of the star could modify the characteristics of
  the key features of the SMBHB system only moderately. 
The high-quality X-ray light curve obtained by S2012 did not 
  only catch the time of the interruption, the recurrence, and the
  re-interruption of the TDE, but also constrains well the
  long durations of the phases of low and high flux. 
  The long duration of the recurrence suggests that
  the resonant perturbation by the secondary BH should not be too
  strong and the merger should not be a major merger with $q \ga 0.3$,
  while the long duration of the interruption implies that the
  perturbation cannot be too weak and the merger should not have an
  extreme mass ratio $q \la 0.01$ irrespective to the detailed
  modeling of the light curve. Therefore, the solutions with $q \sim
  0.08$ are certain. With a moderate small mass ratio $q\sim 0.08$, the
  duration of the recurrent flare should not be much longer than the
  observed one, suggesting that the time of recurrence should not be
  much earlier than the first detection of the recurrent flare and the
  SMBHB orbital period is about the time of recurrence. Therefore, the
  model solutions with $T_{\rm b} \sim 150 \, {\rm days}$ are 
  robust, too. However, we have to notice that the orbital period
  of the SMBHB is the Keplerian period, but the return time of the
  bound stellar plasma elements is determined by the radial
  epicyclic frequency. The Keplerian and epicyclic frequencies are
  equal in Newtonian gravity, but the latter is smaller than the former
  in GR. The differences between the two frequencies become
  significant for tidal disruptions by the SMBH with $M_{\rm BH} \ga
  10^7 M_\sun$ because the tidal radius for a solar-type star is
  $r_{\rm t} \la 5 r_{\rm g} M_7^{-2/3}$. Therefore, for the certain
  orbital period $T_{\rm b}$, the time of the interruption in the
  model light curves is delayed so much that it is significantly later
  than the date of the first upper limit, which, for a SMBHB system
  with circular orbit, is hardly compensated by adjusting the other
  parameters. To significantly shift the time of interruption to an
  earlier date, we need an elliptical orbit for the SMBHB
  system. For the SMBHB system with the given orbital period, the
  moderate eccentricity decreases significantly the pericenter and
  thus the inner boundary of the chaotic shell region, resulting in
  the appearance of the interruption before the first upper limit on
2010 July 7 and after the last detection on 2010 June 30. Therefore,
  we expect that the solution for $M_{\rm BH} = 10^7 M_\sun$ with
  $e_{\rm b} \sim 0.3$ is robust.

%%%%%%%%%%%%
\subsection{Gravitational wave emission}
\label{sec:gw}

Upon final coalescence, SMBHB systems like the one in \sdss are prime
sources for future space-based GW missions like {\it eLISA}. However, 
in its current state of evolution, it would be challenging to detect
the GWs from this system. It would radiate GW emission at a frequency 
$f_{\rm obs} = 2 /T_{\rm b} (1+z) \simeq 0.13\, {\rm {\mu}Hz} \left(T_{\rm
b} /150 \, {\rm days}\right)^{-1}$ in the observer frame, and with a
characteristic strain in an observation of duration $\tau_{\rm obs}$
\begin{equation}
h_{\rm c} = h_{\rm r} \sqrt{f_{\rm obs} \tau_{\rm obs}} ,
\end{equation}
with $h_{\rm r}$ is the strain amplitude at the object rest frequency,
$f_{\rm b} = 2/T_{\rm b}$
\begin{equation}
h_{\rm r} = {8\pi^{2/3} \over 10^{1/2}} {G^{5/3} M^{5/3} \over c^4
  r(z)} f_{\rm b}^{2/3} ,
\end{equation}
where $M = (M_1 M_2)^{3/5} /(M_1+M_2)^{1/5} = M_{\rm BH} q^{3/5} /
(1+q)^{1/5}$ is the ``chirp mass'' of the SMBHB and $r(z)$ is
the comoving distance. The frequency $f_{\rm obs} \simeq 0.13 \,
{\rm {\mu}Hz}$ is in the range observable with PTAs, 
but much smaller than that of {\it eLISA}. For a five-year
PTA observation, the expected characteristic strain of the
binary is
\begin{eqnarray}
h_{\rm c} \approx 1.7\times 10^{-20} \left({T_{\rm b}
  \over 140 \, {\rm days}}\right)^{-7/6} \left({\tau_{\rm obs} \over 5\,
  {\rm yr}}\right)^{1/2} M_6^{5/3} {q_{-1} \over
  (1+q)^{1/3}} \nonumber \\
\approx 6.0 \times 10^{-19} \left({T_{\rm b} \over 150 \, {\rm
  days}}\right)^{-7/6} \left({\tau_{\rm obs} \over 5\,
  {\rm yr}}\right)^{1/2} M_7^{5/3} {(q/0.08) \over (1+q)^{1/3}} ,
\end{eqnarray}
which is about four orders of magnitude smaller than the detection
limit of PTA at the frequency $f_{\rm obs}$
\citep[e.g.,][]{ell12}. If GW radiation was the only
contribution to the orbital shrinkage, the lifetime of the system is
\begin{equation}
\tau_{\rm gw} \simeq 1.0\times 10^8 \, {\rm yr}\, \left({T_{\rm b}
  \over 140\, {\rm days}}\right)^{8/3} M_6^{-5/3} q_{-1}^{-1} f^{-1}
  (1+q)^{1/3}
\label{eq:lfgw}
\end{equation}
\citep{pet63}, where $f$ is a function of the eccentricity $e_{\rm b}$
\begin{equation}
  f = \left(1 + {73 \over 24}e_{\rm b}^2 + {37 \over 96} e_{\rm
  b}^4\right) \left(1-e_{\rm b}^2\right)^{-7/2} .
\end{equation}
For a SMBHB system with $M_{\rm BH} = 10^7 M_\sun$ and typical
parameter values $T_{\rm b} = 150\, {\rm days}$, $q=0.08$, and
$e_{\rm b}= 0.3$, the life time is $\tau_{\rm gw} \simeq 1.9
\times 10^6 \, {\rm yr}$, while for a SMBHB system with $M_{\rm BH} =
10^6 M_\sun$ and the parameter values $T_{\rm b} = 140\, {\rm
days}$, $q=0.1$ and $e_{\rm b}= 0$, the life time is longer, with 
$\tau_{\rm gw} \simeq 1.0\times 10^8 \, {\rm yr}$. However, a
longer life time $\tau_{\rm gw}$ does not necessarily imply that the
$10^6\msun$ fit is favored, nor that TDE light curve
searches for SMBHBs are biased toward systems with larger 
separations and smaller SMBH masses for a given observed binary 
period. This is because of the biases in event detection in 
any (X-ray, transient) survey, which preferentially selects for the
most luminous, most frequent events. In particular, X-ray surveys are
biased toward SMBHs of higher mass $M_{\rm BH} \approx 10^7 \msun$
because of their higher peak luminosities. Further, the 
TDE rate $N_{\rm TDE}$ is a function of $a_{\rm b}$. 
A preliminary estimation suggests that $N_{\rm TDE}$ is a complicated
function of $a_{\rm b}$ and does not change significantly for $10^{-3}
\, {\rm pc} \la a_{\rm b} \la  1 \, {\rm pc}$ \citep{che11,liu12}.
Thus, for
a given observed $T_{\rm b}$ in an X-ray transient survey, a SMBHB
system with higher SMBH mass $M_{\rm BH} \sim 10^7 \msun$ and smaller
separation within the range of $10^{-3} \, {\rm pc} \la a_{\rm b} \la
1 \, {\rm pc}$ would be likely favored.

\subsection{Frequency of SMBHBs among known TDEs}
\label{sec:freq}

Approximately 20 to 25 TDEs and candidates have been identified from
the observations \citep[see][for a recent review]{kom12}, including the
X-rays \citep[e.g.,][]{kom99},
UV \citep{gez08}, optical and emission lines \citep{kom08b,van11}, and
gamma-rays \citep[e.g.,][]{bur11,blo11,cen12}. We have inspected these
light curves, with the exception of the two jetted TDEs, in order
to search for similar features as seen in
SDSSJ1201+30. None are clearly present. This might have several
reasons. It could be due to the gaps in the light curves of 
the TDEs (and so escaped detection). Alternatively, the orbital timescales
of the SMBHBs can be longer or shorter, and so their effects would
be undetectable on the timescales observed so far. Currently,
this then makes SDSS J1201+30 the only good SMBHB candidate among the
known TDEs.

In summary, we conclude that the SMBHB model for SDSS J1201+30
is a viable model that naturally explains the abrupt dips
and recoveries of the X-ray light curve of the event. Future sky
surveys are expected to detect TDEs in the thousands. Analysis
of their light curves will then provide a powerful new tool of
searching for SMBHBs in otherwise non-active galaxies.

\acknowledgments

We are grateful to Peter Berczik, Xian Chen, Zoltan Haiman, Luis Ho,
Richard Saxton, Rainer Spurzem, and Takamitsu Tanaka for helpful
discussions. We would like to thank the anonymous referee for the
  very useful comments and suggestions. 
This work is supported by the National Natural 
Science Foundation of China (NSFC11073002 and NSFC11303039). F.K.L.
thanks the open funding (No. Y3KF281CJ1) from Key Laboratory of
Frontiers in Theoretical Physics, Institute of Theoretical Physics, of 
CAS for financial support. S.L. acknowledges support by NAOC CAS
through the Silk Road Project (grant No. 2009S1-5 for R. Spurzem)
and the ``Qianren'' (Thousand Talents Plan) Project for R. Spurzem.
S.K. would like to thank the Kavli
Institute for Theoretical Physics (KITP) for their hospitality and
support during the program on ``A universe of black holes.'' This
research was supported in part by the National Science Foundation
under grant No. NSF PHY11-25915. The computations have been done on
the Laohu supercomputer at the Center of Information and
Computing at National Astronomical Observatories, Chinese Academy of
Sciences, funded by Ministry of Finance of People's Republic of China
under the grant $ZDYZ2008-2$.

\end{document}